\documentclass[prx,twocolumn,showpacs,amsmath,amssymb,superscriptaddress,bibnotes]{revtex4-2}

\usepackage{graphicx}
\usepackage{color}
\usepackage{braket}
\usepackage{amsmath}
\usepackage{amsfonts}
\usepackage[caption=false]{subfig}
\usepackage{natbib}
\usepackage{bbold}
\usepackage{verbatim}
\usepackage{amsmath,amssymb,mathrsfs}
\usepackage{bbm}
\usepackage[dvipsnames]{xcolor}

\newcommand{\new}[1]{{\color{black}{#1}}}
\newcommand{\cw}[1]{{\color{black}{#1}}}

\allowdisplaybreaks 
\date{\today}

\begin{document}

\title{Intertwined Superconductivity and Magnetism from Repulsive Interactions \\ in Kondo Bilayers}

\author{Clara S. Weber}
\email[Email: ]{clara.weber@rwth-aachen.de}
\affiliation{Institut f\"ur Theorie der Statistischen Physik, RWTH Aachen, 
	52056 Aachen, Germany and JARA - Fundamentals of Future Information Technology}
\affiliation{Department of Physics and Astronomy, University of Pennsylvania, Philadelphia, Pennsylvania 19104, USA}
\author{Dominik Kiese}
\affiliation{Center for Computational Quantum Physics, Flatiron Institute, 162 5th Ave, New York, NY 10010}
\author{Dante M. Kennes}
\affiliation{Institut f\"ur Theorie der Statistischen Physik, RWTH Aachen, 
	52056 Aachen, Germany and JARA - Fundamentals of Future Information Technology}
\affiliation{Max Planck Institute for the Structure and Dynamics of Matter, Center for Free Electron Laser Science, 22761 Hamburg, Germany}
\author{Martin Claassen}
\email[Email: ]{claassen@sas.upenn.edu}
\affiliation{Department of Physics and Astronomy, University of Pennsylvania, Philadelphia, Pennsylvania 19104, USA}
\affiliation{Center for Computational Quantum Physics, Flatiron Institute, 162 5th Ave, New York, NY 10010}

\begin{abstract}

While superconductors are conventionally established by attractive interactions, higher-temperature mechanisms for emergent electronic pairing from strong repulsive electron-electron interactions remain under considerable scrutiny.
Here, we establish a strong-coupling mechanism for intertwined superconductivity and magnetic order from purely repulsive interactions in a Kondo-like bilayer system, composed of a two-dimensional Mott insulator
coupled to a layer of weakly-interacting itinerant electrons. Combining large scale DMRG and Monte Carlo simulations, we find that superconductivity persists and coexists with
magnetism over a wide range of interlayer couplings. We classify the resulting rich phase diagram and find 2-rung antiferromagnetic and 4-rung antiferromagnetic order in one-dimensional systems along with a phase separation regime, while finding that superconductivity coexists with either antiferromagnetic or ferromagnetic order in two dimensions.
Remarkably, the model permits a rigorous strong-coupling analysis via localized spins coupled to charge-2$e$ bosons through Kugel-Khomskii interactions, capturing the pairing mechanism in the presence of magnetism due to emergent attractive interactions. Our numerical analysis reveals that pairing remains robust well beyond the strong-coupling regime, establishing a new mechanism for superconductivity in coupled weakly- and strongly-interacting electron systems, relevant for infinite-layer nickelates and superconductivity in moir\'e multilayer heterostructures.

\end{abstract}

\maketitle
\section{Introduction}

\begin{figure}[t]
    \centering
    \includegraphics[width=1.0\columnwidth]{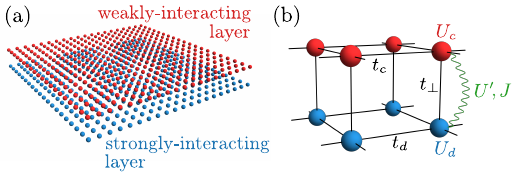}
    \caption{(a) Schematic representation of the bilayer system which is composed of a weakly-interacting layer (shown in red) and a strongly-interacting Mott-insulating layer (shown in blue).  Each of the layers has a square lattice structure.
    (b) Depiction of different terms of the Hamiltonian \eqref{eq:ham_complete}. It contains the intralayer hoppings $t_d$ and $t_c$, the interlayer hopping $t_\perp$, local Hubbard interactions $U_d$ and $U_c$, as well as an interlayer repulsion $U^\prime$ and a Hund's coupling $J$.}
    \label{fig:model}
\end{figure}

The emergence of unconventional high-temperature superconductivity from strong repulsive electronic interactions has captivated condensed matter research since the discovery of the cuprates \cite{dagotto_94,sigrist_ueda_rmp_91,norman_science_11,stewart_aip_17}. In contrast to phonon-mediated conventional superconductors described by BCS theory \cite{bardeen_etal_pr_57}, electronic pairing from Coulomb repulsion fundamentally relies on an emergent effective attraction at low energy scales. For weak electronic repulsion with respect to the single-particle bandwidth, perturbative and renormalization group (RG) approaches are well-understood to yield pairing instabilities such as Kohn-Luttinger superconductivity \cite{kohn_luttinger_prl_65, Maiti_2013, kagan_jetp_14, cao_etal_prl_20, cea_etal_prb_22, wagner_etal_arxiv_23}, spin-fluctuation exchange pairing near spin-density-wave order \cite{scalapino_86, noack_etal_pmb_96, 
scalapino_jltp_99, moriya_pja_06, chang_etal_epjb_20} or Hund's interactions exceeding density repulsion in multi-orbital systems \cite{vafek_chubokov_prl_17, cheung_agterberg_prb_19}. 
In contrast, elucidating the emergence of higher-temperature superconductivity from \textit{strong} repulsive interactions in systems such as doped cuprates, nickelates, or small-twist-angle moir\'e heterostructures remains challenging, as conventional perturbative approaches cease to apply. Here, the undoped parent compounds are Mott insulators, and superconductivity emerges upon doping and in a competition with charge and magnetic order.

Much of the current understanding of the strongly-correlated electron problem rests on numerical approaches such as the density matrix renormalization group (DMRG), quantum Monte Carlo (QMC), and dynamical mean field theory calculations (DMFT). In the past decades, numerous studies of one-dimensional and ladder Hubbard models using DMRG and tensor networks 
\cw{\cite{karakonstankis_etal_prb_11, ehlers_eal_prb_17, jiang_devereaux_science_19, jiang_devereaux_fron_23, jiang_prb_23}} 
have revealed that an understanding of unconventional superconductors requires disentangling a rich interplay between superconductivity and charge or magnetic orders at strong coupling. Effective inverted Hund's coupling in multi-orbital Hubbard models were investigated using DMFT, as a mechanism for unconventional superconductivity in fullerides \cite{georges_etal_arcmp_13, nomura_etal_sa_15, pavarini_book_17, fanfarillo_etal_prl_20}, and the effects of  self-doping and Kondo coupling on antiferromagnetic correlations in interacting bilayers were investigated in QMC \cite{liu_npj_qmat_24}.
Complementarily, explicit strong-coupling theories of emergent electronic attraction from strong repulsive interactions have been constructed for dimerized Hubbard models \cw{\cite{tsai_kivelson_prb_06}}
, spinless fermions \cw{\cite{slagle_kim_scipost_19}} 
, and multiband excitonic models \cite{crepel_fu_scad_21, crepel_etal_prb_22, crepel_fu_proc_22, he_etal_prr_23}, permitting new insights into superconductivity from strong Coulomb repulsion alone. However, these theories by design do not admit scenarios of intertwined or concomitant charge or magnetic orders that are expected to be relevant for high-$T_c$ systems. 
\cw{Examples include infinite layer nickelate superconductors, where superconductivity coexist with short-range magnetic correlations of the Ni $3d$ orbitals \cite{zhang_etal_prb_20, yang_zhang_fron_22, chen_etal_prb_22, chen_etal_fron_22, fowlie_etal_natphys_22, nomura_arita_iop_22}, and superconductivity in magic-angle twisted trilayer graphene \cite{park_etal_nature_21}, where strong electronic pairing emerges from interacting flat bands coupled to a dispersive itinerant Dirac metal. Additionally, the interplay between superconductivity and magnetism plays a decisive role in heavy fermion superconductors on Kondo lattices, where magnetic moments and conduction electrons coexist \cite{bodensiek_etal_jop_10, lynn_etal_fron_23}.} This immediately raises a two-fold question: (1) fundamentally, whether the competition and coexistence of superconductivity, magnetic and charge order at \textit{strong} coupling can be disentangled in a controlled setting, and (2) whether such settings support a new route for higher-temperature superconductivity that can be realized in physical models and materials, and away from strong coupling?

In this \cw{work}, we answer these questions affirmatively, and propose a novel mechanism for \textit{intertwined} superconductivity and magnetic order in Kondo bilayers from strong and purely repulsive interactions. Our work rests on the tantalizing observation that the usual effective Kondo exchange coupling $J_K ~ \hat{\mathbf{s}} \cdot \hat{\mathbf{S}}_i$ between an impurity spin $\hat{\mathbf{s}}$ and a neighboring metallic site $i$ with $\hat{\mathbf{S}} = \hat{\Psi}^\dag_{i} \boldsymbol{\sigma} \hat{\Psi}_{i}$, which arises from second order co-tunnelling processes $\sim t^2/U$ \cite{schrieffer_66}, generically has a subleading \textit{attractive} interaction $U_{\rm eff} \hat{\Psi}^\dag_{i\uparrow} \hat{\Psi}_{i\uparrow} \hat{\Psi}^\dag_{i\downarrow} \hat{\Psi}_{i\downarrow}$ for electrons in the metal which emerges to \textit{fourth} order $U_{\rm eff} \sim t^4/U^3$ in hybridization $t$ and impurity interactions $U$. While a lattice of magnetic moments interacting with a proximal metal will ordinarily favor a Kondo insulator at low temperatures \cite{coleman,zhao_etal_nature_23}, in this work we show that the inclusion of longer-ranged repulsive Coulomb interactions between the magnetically-ordered Mott insulator and neighboring metal can readily tip the balance to instead favor unconventional electronic pairing in the metal. 

The resulting phase diagram comprises a fundamentally new scenario of intertwined magnetic order and superconductivity from strong Coulomb repulsion. It can be motivated already by studying a single rung of an interacting Kondo bilayer, and allows for a rigorous strong-coupling analysis in terms of an effective Kugel-Khomskii-type spin-pairing model of Mott magnetic moments coupled to charge-2$e$ electron pairs in the itinerant metal, which condense to form a superconductor, intertwined with magnetic order. We first present the electronic phase diagram of a two-leg ladder variant of our model using large-scale DMRG calculations as a function of doping and present evidence for superconductivity over a wide parameter space well beyond the validity of the strong-coupling limit. Notably, the strong-coupling superconductor favors pair density wave (PDW) order for a significant part of the phase diagram, while the intertwined magnetic order exhibits a rich phase diagram with 2- and 4-rung antiferromagnetic order. We then study the emergence of intertwined superconductivity and magnetic order in two-dimensional Kondo bilayers using semi-classical Monte Carlo simulations, and observe that electronic pairing persists robustly at finite doping, predicated on the concurrent onset of magnetic order in the insulating layer. Our results establish a new mechanism for high-temperature superconductivity that is relevant to infinite-layer nickelates as well as moir\'e heterostructures such as twisted trilayer graphene and twisted transition-metal dichalcogenides.

\section{Kondo Bilayers}

We consider a heterostructure composed of a Mott-insulating layer with strong Hubbard interactions coupled to a weakly-interacting metallic layer, depicted in Fig.~\ref{fig:model}. Its Hamiltonian 
\begin{align}
\hat{H} &= -\sum_{\langle i j\rangle \sigma}\left(t_d \hat{d}_{i \sigma}^{\dagger} \hat{d}_{j \sigma}+t_c \hat{c}_{i \sigma}^{\dagger} \hat{c}_{i \sigma}\right) + \cw{\sum_i}\hat{H}_i^{\text {rung}} - \mu \hat{N}
\label{eq:ham_complete}
\end{align}
describes electrons $\hat{d}_{i\sigma}$ ($\hat{c}_{i\sigma}$) in the Mott insulating (metallic) layer that hop with amplitudes $t_d$ ($t_c$). Interactions and interlayer hybridization can be usefully decomposed into individual rungs $\hat{H}_i^{\text{rung}}$ of adjacent strongly-interacting and a weakly-interacting sites $i$ in the Mott and metallic layer, respectively. Each rung 
\begin{align}
	\hat{H}_i^{\text {rung}} &= U_c (\hat{n}^c_{i \uparrow} - \tfrac{1}{2}) (\hat{n}^c_{i \downarrow} - \tfrac{1}{2}) + U_d (\hat{n}^d_{i \uparrow} - \tfrac{1}{2}) (\hat{n}^d_{i \downarrow} - \tfrac{1}{2})  \notag\\
	&+ \sum_{\sigma\sigma'} (U' - J \delta_{\sigma,\sigma'}) \hat{n}_{i\sigma}^c \hat{n}_{i\sigma'}^d - t_\perp \left( \hat{c}_{i \sigma}^{\dagger} \hat{d}_{i \sigma}+\text {h.c.} \right) \notag\\
	&+ J \left(\hat{d}_{i \uparrow}^{\dagger} \hat{d}_{i \downarrow}^{\dagger} \hat{c}_{i \downarrow} \hat{c}_{i \uparrow} - \hat{d}_{i \uparrow}^{\dagger} \hat{d}_{i \downarrow} \hat{c}_{i \downarrow}^{\dagger} \hat{c}_{i \uparrow} +\text {h.c.}\right) \label{eq:ham_rung}
\end{align}
is subjected to strong (weak) Hubbard interaction $U_d$ ($U_c$) in the Mott (metallic) layer, while also accounting for inter-layer density $U'$ and Hund's $J$ interactions in Kanamori form. Furthermore interlayer hopping is parameterized by $t_\perp$. The bilayer is half-filled and particle-hole symmetric for a chemical potential $\mu = U' - J/2$.

\section{Strong Coupling Regime}
\begin{figure}[t]
    \centering
    \includegraphics[width=1.0\columnwidth]{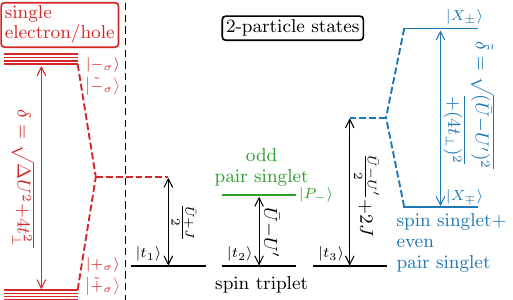}
    \caption{States and energies of a single rung. The two-particle states split up into a spin triplet $\Ket{t_1}, \Ket{t_2}, \Ket{t_3}$ (black), an odd pair singlet $\Ket{P_-}$ (green) and states that are a mixture of spin singlet and even pair singlet $\Ket{X_+}, \Ket{X_-}$(blue). Whether $\Ket{X_+}$ or $\Ket{X_-}$ are lower in energy depends on the exact  parameters. Additionally there are states containing only one electron $\Ket{+_\sigma}, \Ket{-_\sigma}$ or hole $\Ket{\tilde{+}_\sigma}, \Ket{\tilde{-}_\sigma}$ (red). These states are always degenerate due to particle-hole symmetry. If the lower red states have the lowest local energy, this immediately leads to charge-2e fluctuations that could be responsible for superconductivity.  All energies depend on the Hund's coupling $J$, the interlayer hopping $t_\perp$, the interlayer density repulsion $U^\prime$, the average Hubbard repulsion $\bar{U}=(U_c+U_d)/2$, as well as the local interaction difference $\Delta U =(U_d-U_c)/2$. In the whole paper we choose $t_\perp$ to maximize the energy gain of the lower red states.}
    \label{fig:rung_states}
\end{figure}

In the absence of inter-layer interactions, the low-temperature behavior of the bilayer is dominated by Kondo exchange $J_K \sim 4t_\perp^2/U_d$ (assuming $U_d\gg U_c$) between Mott and metallic sites on each rung, which stabilizes a Kondo insulator in the strong coupling limit. To see how competing inter-layer interactions $U'$, $J$ can nudge the metallic layer towards superconductivity, consider first the local excitation spectrum of a single rung $\hat{H}_i^{\text{rung}}$.
Every rung can be occupied by at most four electrons, however empty (zero-electron) and fully occupied (four-electron) states are energetically unfavorable due to strong Hubbard repulsion $U_d$ and can be neglected.
{\color{black} When the bilayer is half-filled ($\mu = U' - J/2$),} the ground state of a single rung would ordinarily be expected to be a spin triplet. However, an imbalance in Hubbard interactions in the two layers $U_d \gg U_c$, combined with interlayer hybridization, can remarkably tip the energetic balance and instead stabilize single-electron or three-electron (single-hole) states as the rung ground state, depicted Fig.~\ref{fig:rung_states}. Single-electron and three-electron states (shown in red) are degenerate at half filling due to particle hole symmetry; both carry a spin $1/2$. The low-energy manifold therefore comprises a spin-$1/2$ magnetic moment (carried by an unpaired electron, primarily though not exclusively localized in the strongly-interacting layer), in the presence or absence of a charge-2$e$ bound (Cooper) pair (that is primarily hosted in the metallic layer) which can condense in a macroscopic system of coupled rungs to form a superconductor.

We now analyze the single-rung spectrum of Fig.~\ref{fig:rung_states} in detail. Interlayer hybridization splits the single electron states into
$\left|+_\sigma\right\rangle=\cos (\theta / 2)\left|\sigma_d, 0_c\right\rangle+\sin (\theta / 2)\left|0_d, \sigma_c\right\rangle$ (bonding; at lower energy) and $\left|-{ }_\sigma\right\rangle=\sin (\theta / 2)\left|\sigma_d, 0_c\right\rangle-\cos (\theta / 2)\left|0_d, \sigma_c\right\rangle$ (antibonding; at higher energy) with a mixing angle of $\tan{\theta} = 2 t_\perp/\Delta U$, {\color{black}$\theta \in [0,\pi/2]$}, where $\Delta U=(U_d-U_c)/2$ and the local spin $\sigma=\uparrow,\downarrow$. The single hole states $\left|\tilde{+}_\sigma\right\rangle$, $\left|\tilde{-}_\sigma\right\rangle$ can be found via particle-hole transformation. The spectrum of two-electron states is more involved: First, spin triplet states (shown in black) are given by $\Ket{t_1}=\Ket{\uparrow_d, \uparrow_d}$, $\Ket{t_2}=\Ket{\downarrow_d, \downarrow_d}$ and $\Ket{t_3}=(\Ket{\uparrow_d, \downarrow_c}+\Ket{\downarrow_d, \uparrow_c})/\sqrt{2}$. The singlet state $\Ket{s}=(\Ket{\uparrow_d, \downarrow_c}-\Ket{\downarrow_d, \uparrow_c})/\sqrt{2}$ hybridizes with the even pair state $\Ket{P_+} = (\Ket{\uparrow\downarrow_d, 0_c}-\Ket{0_d, \uparrow\downarrow_c})/\sqrt{2}$, forming the combinations $\Ket{X_+}=\cos(\phi/2) \Ket{s} + \sin(\phi/2) \Ket{P_+}$
and $\Ket{X_-}=\sin(\phi/2) \Ket{s} - \cos(\phi/2) \Ket{P_+}$ with $\tan(\phi) = 4 t_\perp /(\bar{U}-U^\prime)$, {\color{black}$\phi \in [-\pi/2, \pi/2]$} where $\bar{U}=(U_d+U_c)/2$. These states are marked in blue in Fig.~\ref{fig:rung_states}; their energetic ordering depends on the sign of $\bar{U}-U^\prime$.
Finally, there is an odd pair singlet state (shown in green) $\Ket{P_-}=(\Ket{\uparrow\downarrow_d, 0_c}-\Ket{0_d, \uparrow\downarrow_c})/\sqrt{2}$.
Importantly, if the bound
\begin{align}
\delta>\max \left\{\bar{U}+J, 2 U^{\prime}-\bar{U}+J, U^\prime -3J+\tilde{\delta}\right\}
\label{eq:delta_bound}
\end{align}
is satisfied, then single-electron and single-hole states form the low-energy states on a single rung, with $\delta, \tilde{\delta}$ given in Fig.~\ref{fig:rung_states}, and can be usefully understood as the product of an effective spin-$1/2$ and a charge-$2e$ hardcore boson. Finally, doping with a small finite chemical potential breaks the degeneracy between electron and hole states, acting as an effective chemical potential for the charge-$2e$ boson while keeping the spin-$1/2$ intact. We stress that this scenario occurs for \textit{a priori} purely repulsive interactions with $U^\prime - J>0$. 

In the strong coupling-regime $U_d > U^\prime, J, t_\perp \gg t_c, t_d$ and with only a single layer subjected to strong Hubbard interactions $U_d \gg U_c$, the mixing angle $\theta$ for single-electron and single-hole rung ground states is small, meaning that the magnetic moment is primarily localized in the strongly interacting layer while the Cooper pairs live in the proximal metallic layer. Nevertheless, mixing between both layers is essential as the regime of interaction parameters that stabilizes single-electron and single-hole states as the lowest-energy states per rung shrinks rapidly for small mixing angles.

Suppose now that neighboring rungs are coupled via intra-layer hoppings $t_d$, $t_c$ [Eq. (\ref{eq:ham_complete})] to form a bilayer lattice. At strong coupling, hopping of single electrons is suppressed. However, virtual tunneling processes can act in a two-fold manner to break the degeneracy of the low-energy states per rung. First, superexchange processes lead to an effective exchange coupling between neighboring spins that can be either antiferromagnetic or ferromagnetic, due to interference between virtual tunneling processes $t_d$ and $t_c$ in the strongly-interacting and in the metallic layer, respectively. Second, pairs of electrons can tunnel between neighboring sites via second-order hopping processes, imbuing the charge-2$e$ Cooper pairs with dynamics. However, both effects do not exist in isolation: hopping of Cooper pairs between neighboring rungs must depend on the rungs' magnetic state, and exchange interactions must depend on the charging state and nearest-neighbor hybridization of the charge-2$e$ bosons. Consequently, the emergence of superconductivity is intimately tied to magnetic order in the Mott insulator. The low-temperature physics can be usefully understood as a charge-$2e$ analogon of Kugel-Khomskii spin-orbital interactions, whereby the orbital degree of freedom becomes replaced by local charge-2$e$ fluctuations, and the resulting ground state phase diagram is expected to host a rich competition of competing and intertwined superconducting, spin and charge orders. 

\section{Two-Leg Kondo Ladders}

While we will discuss this strong-coupling analysis below, importantly, we first show that the applicability of this picture extends far beyond the na\"ive strong coupling limit. To this end, we consider a one-dimensional version of the interacting bilayer -- a two-leg ladder with a strongly-interacting and a weakly-interacting leg -- and study its ground state using (i)DMRG as a function of doping.

\begin{figure}[t]
    \centering
    \includegraphics[width=1.0\columnwidth]{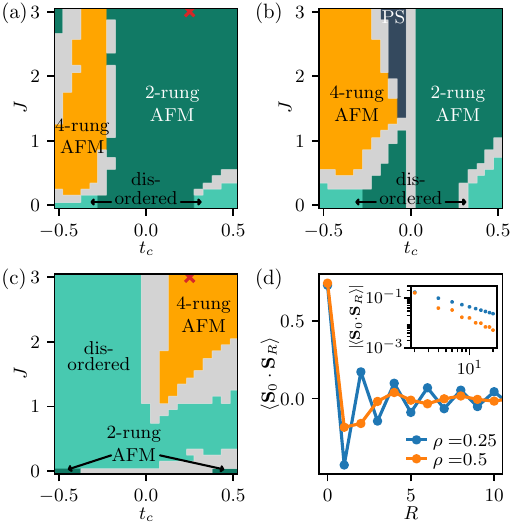}
    \caption{(a),(b) Magnetic phase diagram at quarter filling as a function of the hopping in the metallic layer $t_c$ and the Hund's exchange $J$ for $U_d=1$, $U_c=0$, $U^\prime=0.4$, $t_\perp$ chosen to maximize the local energy gain and (a) $t_d=0.25$ or (b) $t_d=|t_c|$.
In (a) we find a 2-rung AFM behavior for most of the parameters. For $t_c \leq -t_d = -0.25$ we find mostly a 4-rung antiferromagnetic ordering. For very large $J$ and large but negative $t_c$ the system switches back to the 2-rung AFM phase. In the regime of very large $|t_c|$ and small $J$ we find a disordered phase, where no clear magnetic ordering can be observed (either exponential decay of the correlation or no clear peak in the Fourier transform). In the light gray regions we were not able to identify the order (due to a mixture of different phases, an incommensurability of the order) or our data was either not converged or decayed exponentially due to the proximity to a phase boundary.
In b) we find an even richer phase diagram, where we get a phase separation region (PS) for small $t_d=-t_c$ and large $J$ in addition to the phases found in (a).
    (c) Magnetic phase diagram at half filling for the same parameters as (a). The orders have completely changed compared to the results at quarter filling and most parts are disordered now. 
    (d) Magnetic correlation function for different fillings at $t_c=t_d=0.25$ (other parameters as in (a)-(c)) as marked by red crosses in (a) and (c). Inset: Absolute value on a double logarithmic scale {\color{black}for even $R$}. The magnetic ordering changes in dependence on the doping but in both cases we find a power-law decay.}
    \label{fig:magnetic_order}
\end{figure}

We start by characterizing the parent Mott-insulating state, which occurs at \textit{quarter} filling, with a single electron per rung that forms a local moment, and no Cooper pair present in the metallic layer. We work in units of $U_d = 1$ and choose $U_c = 0$ for simplicity. We further constrain, without loss of generality, the following analysis by fixing $t_\perp = \cw{\sqrt{J}/2 \cdot \sqrt{4J + |U_d - 2U^\prime|}}$ 
which maximizes the local energy gain of single-electron/single-hole states with respect to two-electron states per rung. We investigate magnetic order via studying the spin structure factor 
\begin{align}
    S(k) = {\color{black} \frac{1}{L}} \sum_{i,j} e^{-i k (i-j)} \langle \hat{\mathbf{S}}_{i} \cdot \hat{\mathbf{S}}_{j} \rangle
\end{align}
with effective spin operators $[ \hat{\mathbf{S}}_{i} ]^\alpha =  \frac{1}{2} \sum_{b=c, d} [ \hat{b}^\dagger_{i\uparrow} ,\,    \hat{b}^\dagger_{i\downarrow} ] \cdot \sigma^\alpha \cdot [ \hat{b}_{i\uparrow},\, \hat{b}_{i\downarrow} ]^T$ that act on the magnetic moment of the single-electron/single-hole rung states. We seek quasi-long-ranged spin-spin correlations (that decay algebraically in 1D), and identify peaks at $k=0, \pi, \pi/2$ with a ferromagnetic, anti-ferromagnetic (2-rung AFM) and 4-rung anti-ferromagnetic (4-rung AFM) phase, respectively. 
The magnetic phase diagram is shown in Fig.~\ref{fig:magnetic_order}(a) as a function of $t_c$ and $J$ (keeping the other intralayer hopping $t_d = 0.25$ fixed) and exhibits a 2-rung AFM (for $t_c>-\cw{0.25}$) and a 4-rung AFM phase (for $t_c\leq-\cw{0.25}$). For small Hund's couplings and large intralayer hoppings we find a \cw{disordered} regime, beyond the validity of the strong-coupling Mott regime. Additionally, for large J and very negative $t_c$, the system can switch back to the 2-rung AFM phase. Fig.~\ref{fig:magnetic_order}(b) shows similar results for $|t_c|=|t_d|$, where both intralayer hoppings are allowed to become small or large at the same time, with respect to $U_d$. Furthermore, for large Hund's couplings and small $t_c=-t_d$, a phase separation regime appears, with multiple phases degenerate (analyzed using detailed exact diagonalization calculations in the Supplemental Material). Finally, we study the magnetic phase diagram at half filling, depicted in Fig.~\ref{fig:magnetic_order}(c). Here, the weakly-interacting layer becomes half-filled with charge-2$e$ bosons, which has a drastic effect on magnetism: Quasi-long-ranged magnetic order is largely absent, however, a 4-rung AFM phase persists for large $t_c$ and Hund's coupling, \cw{where 2-rung AFM was present at quarter filling}.

In \cw{Fig.~\ref{fig:magnetic_order}(d)} we show for an exemplary parameter set \cw{in this regime} 
how the magnetic ordering \cw{changes in dependence on the filling}. These observations show, that the two layers are not independent but do have an influence on each other. 

\begin{figure*}[t]
    \centering
    \includegraphics[width=1.0\textwidth]{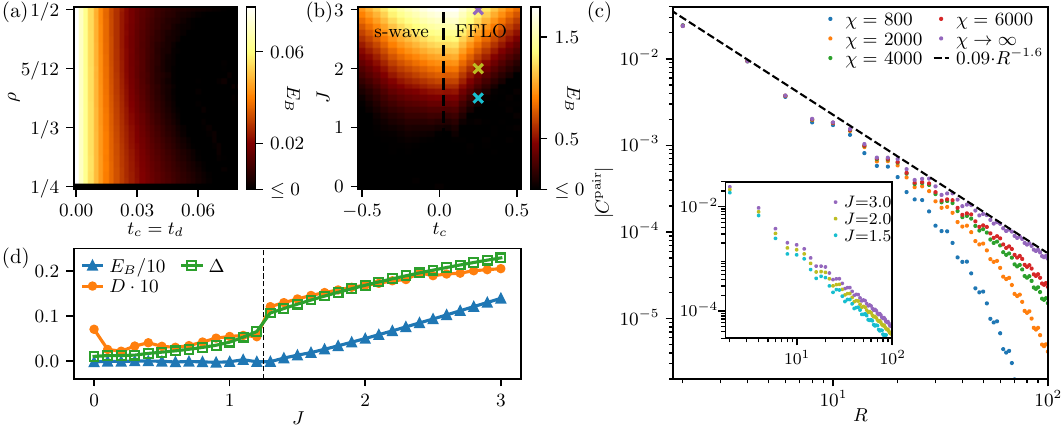}
    \caption{(a) Pair binding energy as a function of $t_c=t_d$ and the filling $\rho$ for $J=0.45$, $U^\prime=0.5$, $U_d=1.0$, $U_c=0.0$ and $t_\perp$ chosen to maximize the local energy gain shown in Fig.~\ref{fig:rung_states}. The pair binding energy is above zero for small intralayer hoppings and it decreases when the intralayer hoppings are increased.
    (b) Pair binding energy as a function of $t_c$ and $J$ for $\rho=3/8$, $t_d=0.25$, and $U^\prime=0.4$ $t_\perp$ chosen to maximize the local energy gain. The other parameters are chosen as in (a). 
    The pair binding energy stays positive in a wide range of the phase diagram being an evidence for superconductivity in this regime. The dashed line shows the phase transition between standard s-wave superconductivity and the exotic FFLO phase. 
    The crosses show at which points the correlation functions in the inset of (c) were calculated.
    (c) Pair correlation function in dependence on the distance $R$ for different bond dimensions and $t_c=t_d=0.25$, $J=3.0$. The other parameters are set according to panel (b). The correlation function gets longer ranged for larger bond dimensions and the extrapolated curve for $D\rightarrow\infty$ follows a power-law as indicated with the dashed line. We only show the results for even $R$ due to the 4-rung periodicity of the correlation function. Inset: Results for $\chi\rightarrow \infty$ are shown for different Hund's exchanges with positive pair binding energy. The parameters correspond to the marked crosses in (b).  
    (d) Pair binding energy $E_B$, pairing strength $D$ with $R_1=5$, $R_2=44$ and order parameter $\Delta$ at $h=0.02$ along $t_c=t_d=0.25$ in (b). The pairing strength and the order parameter both jump between $J=1.2$ and $J=1.3$, where pair binding energy gets finite.}
    \label{fig:pair_correlation}
\end{figure*}

We now investigate the propensity for superconductivity via three complementary metrics: (1) pair correlation functions, (2) the response to an applied pairing bias field, and (3) the pair binding energy
\begin{align}
    E_{\rm B} &= 2E_{N_\uparrow - 1, N_\downarrow} - E_{N_\uparrow -1, N_\downarrow -1} - E_{N_\uparrow, N_\downarrow} \,,
\end{align}
where $E_{N_\uparrow,N_\downarrow}$ is the ground state energy for fixed particle number and spin. 
Fig.~\ref{fig:pair_correlation}(a) shows the pair binding energy as a function of filling and intralayer hoppings $t_c=t_c$, for fixed values of $U^\prime$ and $J$ (keeping $U^\prime-J>0$). Remarkably, the pair binding energy becomes positive for small intralayer hoppings, a first evidence for superconductivity, while becoming zero for larger hopping amplitudes. For $t_c=t_d\rightarrow 0$, the pair binding energy approaches twice the local energy gain of the states with the one electron or hole of a single rung as expected (however superconductivity cannot emerge in this limit, as the individual rungs become uncorrelated).
Complementarily, Fig.~\ref{fig:pair_correlation}(b) shows the pair binding energy at $3/8$-filling as a function of $t_c$ and $J$, and choosing all other parameters as in Figs.~\ref{fig:magnetic_order}(a),(c). We again find a region with positive pair binding energy, which notably persists for smaller Hund's interactions, showing that the superconducting behavior can also be found well outside the validity of the strong coupling analysis.

Since a positive pair binding energy is a necessary but not sufficient criterion for a superconducting phase, we now study the singlet pair correlation function, defined as 
\begin{align}
C^{\mathrm{pair}}_{i,j} &= \frac{1}{4} \left\langle(\hat{d}^\dagger_{i,\uparrow} \hat{d}^\dagger_{i,\downarrow} + \hat{d}_{i,\downarrow} \hat{d}_{i,\uparrow} - \hat{c}^\dagger_{i,\uparrow} \hat{c}^\dagger_{i,\downarrow} - \hat{c}_{i,\downarrow} \hat{c}_{i,\uparrow}) \right.\notag \\
&\cdot \left.(\hat{d}^\dagger_{j,\uparrow} \hat{d}^\dagger_{j,\downarrow} + \hat{d}_{j,\downarrow} \hat{d}_{j,\uparrow} - \hat{c}^\dagger_{j,\uparrow} \hat{c}^\dagger_{j,\downarrow} - \hat{c}_{j,\downarrow} \hat{c}_{j,\uparrow})\right \rangle \,.
\end{align} 
Fig.~\ref{fig:pair_correlation}(c) shows that the pair correlation function within the region of positive pair binding energy. We find that the pair correlation function follows a power-law decay, establishing further evidence for a quasi-long-range-ordered superconducting state in the one-dimensional limit. As matrix product states cannot capture algebraically decaying correlations at finite bond dimension,  we carefully extrapolate the results to infinite bond dimension, described in the Supplemental Material. The inset of Fig.~\ref{fig:pair_correlation}(c) furthermore depicts results with infinite bond dimension for other values of the interlayer Hund's coupling which place the superconductor closer to the phase transition to a metallic normal state, corroborating the identification of a broad regime of superconductivity. Those decay with a power-law as well.
Notably, the pair correlation function furthermore can exhibit periodic oscillations [see Supplementary Material], which we identify as a PDW or Fulde–Ferrell–Larkin–Ovchinnikov (FFLO) phase [Fig. \ref{fig:pair_correlation}(b)]. At $\rho=3/8$, we find a standard s-wave for $t_c/t_d \leq 0$ and an FFLO state for $t_c/t_d > 0$, shown in Fig.~\ref{fig:pair_correlation}(b). We note that the FFLO pairing momentum changes with filling, suggesting an incommensurate FFLO phase upon doping. However, a crisp identification of possibly incommensurate FFLO orders is numerically challenging for arbitrary interaction parameter sets, as discussed in the Supplemental Material. 

\cw{Since we utilize a one-dimensional system for the calculation, the correlation function can still follow a power-law scaling outside the superconducting regime as multiple correlation functions can show a power-law behavior at the same time \cite{giamarchi_book_03}. Therefore, we refrain from comparing the shape of the scaling to define the phase transition. Instead, we define the pairing strength \cite{patel_etal_prb_17, nocera_etal_prb_18} $D=\sum_{R_1}^{R_2} |C^{\rm{pair}}(R)|$ and show the results in Fig.~\ref{fig:pair_correlation}(d) for a fixed line of the phase diagram (b). We observe a jump in the pairing strength at the same point where the pair binding energy gets finite. It is verified that the choice of range $R_1 < R < R_2$ for defining the pairing strength does not influence the qualitative behavior. }

Finally, we calculate the response to a pairing bias field \cite{qin_etal_prx} as a third evidence for superconductivity. To correctly seed the phase, it is crucial that the bias field has the periodicity as expected from the pair correlation function. We therefore add a bias field $\hat{H} = (h/2) \sum_j ( \hat{\Delta}_{j}^{\textrm{LO}}+ \textrm{h.c.} )$
with $\hat{\Delta}_j^{\textrm{LO}}=\cos(j \cdot k) (\hat{d}_{j, \downarrow} \hat{d}_{j, \uparrow} - \hat{c}_{j, \downarrow} \hat{c}_{j, \uparrow})$ to the system, which corresponds to a Larkin-Ovchinikov (LO) state that usually yields a lower energy than the corresponding Fulde-Ferrell (FF) state with $\hat{\Delta}_j^{\textrm{FF}}=\exp(j \cdot k) (\hat{d}_{j, \downarrow} \hat{d}_{j, \uparrow} - \hat{c}_{j, \downarrow} \hat{c}_{j, \uparrow})$ for systems without spin orbit coupling \cite{kinnunen_etal_rpp_18, baarsma_jmo_16, baarsma_stoof_pra_13, mora_combescot_05, yoshida_yip_pra_07}. 
The response to the pairing field is defined as $\Delta=|\braket{\hat{\Delta}_j^{\textrm{LO}}}|$ for every $j$ with $|\cos(k\cdot j)|=1$, \cw{ensuring that we measure the amplitude of the order parameter.}
We fix the chemical potential so that the desired particle number is restored without the pairing field. Fig.~\ref{fig:pair_correlation}(d) shows the results 
for a fixed pairing field strength $h = 0.02$ and $k=\pi/2$, to match the periodicity of the correlation function [see Supplemental Material]. The pair binding energy, pairing strength, and superconducting order parameter all show a jump {\color{black}or get positive} for a finite $J$ that agrees with the analyses above. In combination, all three metrics for superconductivity agree and indicate a broad region of superconductivity that emerges from strong repulsive interactions.

\section{Effective Strong Coupling Model}

Motivated by the observed robustness of strong-coupling superconductivity well beyond the strict strong-coupling limit, we now return to studying the coupled dynamics of effective per-rung effective $S=1/2$ magnetic moments and charge-2$e$ bosons dynamics in detail. To this end, we perform a strong-coupling expansion to second order in intralayer hopping and replace the charge-2$e$ hardcore bosons by pseudo-spin operators $\hat{\boldsymbol{\tau}}_i$, arriving an effective \textit{spin-pairing} Hamiltonian
\begin{align}
\hat{H} &= \sum_{\langle ij \rangle} \new{J_s} \, \hat{\mathbf{S}}_i \cdot \hat{\mathbf{S}}_j \notag \\ 
&+ \sum_{\langle ij \rangle} (K_{xy} + 4 W_{xy} \, \hat{\mathbf{S}}_i \cdot \hat{\mathbf{S}}_j) \, (\hat{\tau}_i^x \hat{\tau}_j^x + \hat{\tau}_i^y \hat{\tau}_j^y) \notag \\ &+ \sum_{\langle ij \rangle} (K_{z} + 4 W_{z} \, \hat{\mathbf{S}}_i \cdot \hat{\mathbf{S}}_j) \, \hat{\tau}_i^z \hat{\tau}_j^z ~-~ \mu \sum_i \tau_i^z \, . 
\label{eq:eff_ham}
\end{align}
This Hamiltonian describes the dynamics of charge-$2e$ hardcore bosons with $U(1)$ charge conservation symmetry coupled to physical $S=\frac{1}{2}$-spins per rung with $SU(2)$ rotation symmetry. Neighboring physical spins interact via exchange interactions \new{$J_s$}. Conversely, neighboring Cooper-pair pseudospins are subjected to XXZ interactions, with XY coupling $K_{xy}$ (equivalently corresponding to hardcore bosons hopping on a lattice) and Ising coupling $K_{z}$ (equivalently describing nearest-neighbor interactions between the charge-2$e$ bosons). Both couplings get renormalized due to the magnetic background (via $W_{xy}$ and $W_z$). Furthermore, a chemical potential acts as an effective magnetic field \new{in the $z$-direction} for the charge-2$e$ pseudospins. The parametric dependence of the effective model parameters \new{for an exemplary set of electronic hoppings and interactions} is depicted in Fig.~\ref{fig:exchange_couplings}. Analytic expressions for the effective exchange parameters are provided in the Supplementary Material. In two dimensions, replacing the physical spin couplings with their mean value readily results in an effective XXZ model for charge-2$e$ bosons that is know to host superfluidity \cite{schmid_etal_prl_02}, corresponding to layer-odd singlet superconductivity in our system.

\section{Emergent Superconductivity in Two-Dimensional Bilayers}

\begin{figure}
    \centering
    \includegraphics[width = 0.9\columnwidth]{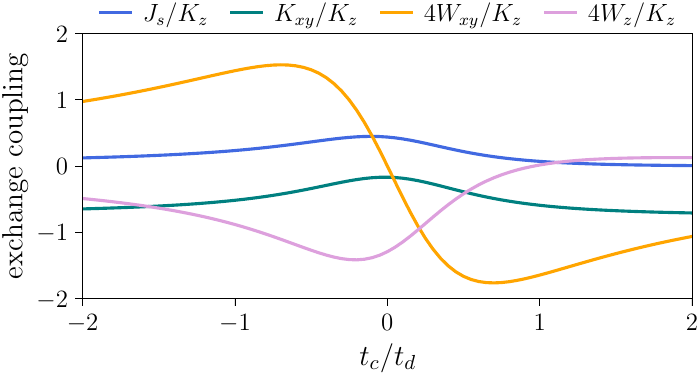}
    \caption{Variation of the spin/pseudospin exchange couplings with the metallic interlayer hopping $t_c$.~\new{The parameters of the bilayer Hubbard-Kanamori model in Eq.~\eqref{eq:ham_complete} are $U_c / U_d = 0, J / U_d = 0.45, U' / U_d = 0.5$ with $t_{\perp} / U_d = 0.45$.} Due to the absence of a single crossing point for all lines, the model does not host an SU(4)-symmetric point.}
    \label{fig:exchange_couplings}
\end{figure}

\begin{figure*}
    \centering
    \includegraphics[width = \linewidth]{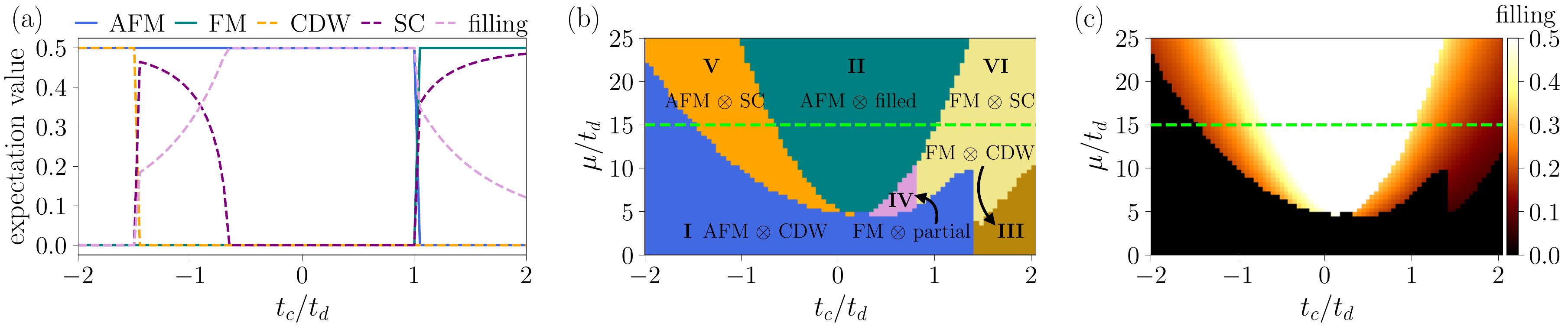}
    \caption{Monte Carlo results for the strong-coupling Kugel-Khomskii Hamiltonian. Simulations are performed on a 24 $\times$ 24 square lattice subject to periodic boundary conditions.~(a) Measured spin and pseudospin expectation values for fixed chemical potential $\mu / t_d = 15$ as indicated by the dashed lines in (b) \& (c).~(b) Phase diagram of the Kugel-Khomskii model as a function of $t_c$ and $\mu$. We find two extended superconducting phases with antiferromagnetic or ferromagnetic order in the insulating layer (phases V and VI, respectively).~(c) Filling of the metallic layer above half-filling for the parameters in (b).}
    \label{fig:QMC_summary}
\end{figure*}

We therefore investigate the intriguing possibility of an extended region of superconductivity in the full two-dimensional effective strong-coupling spin-pairing model \eqref{eq:eff_ham} by calculating its semi-classical phase diagram as a function of intralayer metallic hopping $t_c$ and chemical potential $\mu$ (see Fig.~\ref{fig:exchange_couplings} for the dependence of the exchange couplings on $t_c$). Utilizing the Monte Carlo algorithm from Ref.~\cite{Gresista_2023}, that is, by representing the ground state as a product state and minimizing the energy in the fundamental representation of $\mathfrak{su}$(4) spin-pseudospin operators, we find a rich phase diagram of intertwined magnetic and charge or superconducting orders, shown in Fig.~\ref{fig:QMC_summary}(b). At small chemical potential, the metallic layer is fully charge-polarized with hardcore bosons ordering in a checkerboard arrangement, indicated by the staggered magnetization $\langle \tau^z \rangle$ approaching its maximum value $\tfrac{1}{2}$ (in our representation of the strong-coupling model, all generators have eigenvalues $0$ and $\pm \tfrac{1}{2}$). The insulating layer, on the other hand, is mostly N\'eel ordered (phase I) with a sliver of ferromagnetic correlation at large positive $t_c$ (phase III). This situation changes dramatically upon doping (amounting to increasing the effective magnetic field felt by the charge-2$e$ pseudospins; see Figs.~\ref{fig:QMC_summary}(a) \& (b)). For weak dispersion ($|t_c| \ll 1$), the system quickly transitions from half filling to a trivial insulating phase with a completely filled metallic layer (phase II). However, for finite interlayer hopping $|t_c|$, the weakly-interacting layer instead first becomes partially depleted; the resulting strong charge-2$e$ fluctuations drive a $q = 0$ U(1)-breaking ground state (phases V and VI) and, thus, superconductivity. Remarkably, the observed Cooper pairing is remarkably robust to the type of magnetic order present in the insulating layer, which can either be antiferromagnetic (for phase V) or ferromagnetic (for phase VI), and remains stable over a wide range of parameters. We also find a small intermediate regime (labeled IV in Fig.~\ref{fig:QMC_summary}(b)), where condensation of hardcore bosons is suppressed even though the metallic layer is partially-filled (cf. Fig.~\ref{fig:QMC_summary}(c)). Since the insulating layer is not fully spin-polarized [see Fig.~10(d) in the supplemental material], we speculate that magnetic order therein is a crucial ingredient to stimulate a sufficiently strong attraction between electrons in the metallic layer.

\section{Discussion}

This work establishes a new mechanism for intertwined superconductivity and magnetic order from strong and purely repulsive interactions in Kondo-like bilayer systems. At strong coupling, the interplay of superconductivity and magnetism is captured concisely via an effective theory of tightly-bound Cooper pairs interacting with local magnetic moments, in analogy to Kugel-Khomskii models of spin-orbital order; however, the superconducting phase is found numerically to extend well beyond the strong coupling regime, while coexisting with magnetic order.

Experimentally, twisted or lattice-mismatched moir\'e multilayers of graphene or transition-metal dichalchogenides constitute an attractive target for realization. While originally observed in magic-angle twisted bilayer graphene \cite{cao_tbg_2018}, superconductivity in moir\'e materials has meanwhile been reported in mirror-symmetric twisted graphene trilayers \cite{park_etal_nature_21} as well as twisted bilayer WSe$_2$ \cite{xia_2024}. Near the magic angle, twisted trilayer graphene again forms an almost dispersionless strongly-interacting band, which however hybridizes with copies of pristine dispersive monolayer Dirac cones \cite{mora_etal_prl_19, khalaf_etal_prb_19, carr_etal_nl_20, lei_etal_prb_21, calugaru_etal_prb_21, shin_etal_prb_21, park_etal_nature_21, hao_etal_science_21, christos_etal_prx_22, fischer_etal_nat_22}. The latter can be viewed as a proximal metallic layer, while remaining strongly Coulomb-coupled to the flat band as both flat-band and metallic states originate from orbitals in the same trilayer. Hybridization can be controlled via displacement fields, while the relative strengths of intra- and interband interactions depends on the screening environment. Similarly, moir\'e heterostructures of transition-metal dichalcogenides such as MoTe$_2$/WSe$_2$ heterobilayers have been shown to realize synthetic Kondo lattices \cite{zhao_etal_nature_23}, constituting a highly-tunable platform to search for intertwined superconductivity and magnetism. Furthermore, our mechanism could play a decisive role in infinite-layer nickelates. Here, the nickel $3d$ orbitals are subjected to strong Coulomb repulsion while hybridizing with itinerant rare-earth 5$d$ states; while only short-range magnetic correlations are present at half filling \cite{zhang_etal_prb_20, yang_zhang_fron_22, chen_etal_prb_22, chen_etal_fron_22, fowlie_etal_natphys_22, nomura_arita_iop_22}, superconductivity again emerges upon doping.

Finally, it will be interesting to study the roles of frustration and spin-orbit coupling in altering the interplay of superconductivity and magnetism. Here, in analogy to spin-orbital Kugel-Khomskii models, a wealth of intriguing possibilities ranging from concomitant superconductivity and frustrated magnetic order to superconducting versions of spin-orbital quantum liquid states \cite{PhysRevB.80.064413,PhysRevX.2.041013} warrants further investigation.

\section{Acknowledgement}
M.C. acknowledges support from the NSF under Grant No. DMR-2132591, as well as support through an Alfred P. Sloan foundation fellowship.
C.S.W. and D.M.K are supported by the Deutsche Forschungsgemeinschaft via RTG 1995.
D.M.K acknowledges support from the Deutsche Forschungsgemeinschaft (DFG, German Research Foundation) under Germany’s Excellence Strategy - Cluster of Excellence Matter and Light for Quantum Computing (ML4Q) EXC 2004/1 - 390534769 and within the Priority Program SPP 2244 ``2DMP'' - 443273985. The Flatiron Institute is a division of the Simons Foundation. We acknowledge support from the Max Planck-New York City Center for Non-Equilibrium Quantum Phenomena. Simulations were performed with computing resources granted by the University of Pennsylvania, the Simons Foundation Flatiron Institute, and by RWTH Aachen University under projects rwth0752, rwth0841 and rwth 1258. The authors gratefully acknowledge computing time on the supercomputer JURECA\cite{JURECA} at Forschungszentrum Jülich under grant no. enhancerg. DMRG calculations were performed using the TeNPy Library \cite{tenpy}.

\appendix

\section{Extrapolation of DMRG simulations for Infinite Bond Dimension}
\label{sec:extrapolation}

In this appendix we explain in detail how the extrapolation to infinite bond dimension was performed and argue that this method leads to accurate results. Every matrix product state (MPS) has a finite correlation length by construction. Conversely, algebraically-decaying correlations lead to an infinite correlation length, and hence are not representable via an MPS with finite bond dimension. In order to avoid convergence problems, one can extrapolate observables, in our case the pair correlation function, to infinite bond dimensions.

\begin{figure}[t]
    \centering
    \includegraphics[width=1.0\columnwidth]{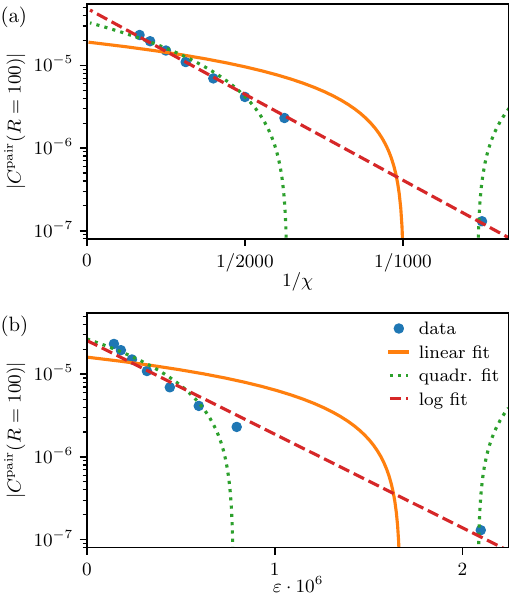}
    \caption{Pair correlation function at $R=100$ and comparison of fits to the inverse band dimension and maximal truncation error for $t_A=t_B=0.25$, $J=3.0$, $V=0.4$, $U_d=1.0$, $U_c=0.0$ and $t_\perp$ chosen to maximize the local energy gain via charge disproportion. The results are shown in dependence on (a) the inverse bond dimension $1/\chi$ and (b) the maximal truncation error $\varepsilon$. The most accurate fit is achieved via a logarithmic ansatz ($\ln{y} = a\cdot x +b$) as a function of $1/\chi$, \cw{which describes the data better that the other fits in dependence on $1/\chi$ or all fits in dependence on $\varepsilon$}.
    Therefore we use this procedure for the results with infinite bond dimension in the main text.}
    \label{fig:extrapolation}
\end{figure}

It is still an open question how to best perform these kind of extrapolations. A commonly used technique is to fit a function in dependence of either the inverse bond dimension, the variance or the truncation error \cite{dolfi_etal_prb_15, hubig_etal_prb_18}. Additionally, it is not obvious which function should be employed to achieve an optimal fit.
Since the calculations are already well converged at small distances we are mainly interested in finding a good approximation for large distances. In Fig.~\ref{fig:extrapolation} we compare the different extrapolation methods for $R=100$ and parameters of Fig.~4(c) (main panel) of the main text. We apply a linear, a quadratic and a logarithmic function in dependence on (a) the inverse bond dimension $1/\chi$ and (b) the maximal truncation error $\varepsilon$. We find that the results are most reasonably captured via a logarithmic fit as a function of $1/\chi$. \cw{All other fits in dependence on $1/\chi$, as well as all fits in dependence on maximal truncation error $\varepsilon$ provide worse results.} 
Therefore, this method is used in the main text whenever pair correlation functions for infinite bond dimensions are shown. Furthermore, we use this technique for extrapolating spin
correlation functions to infinite bond dimensions \cw{as well}.

\section{Filling Dependent Pair Binding Energy}

\begin{figure}[t]
    \centering
    \includegraphics[width=1.0\columnwidth]{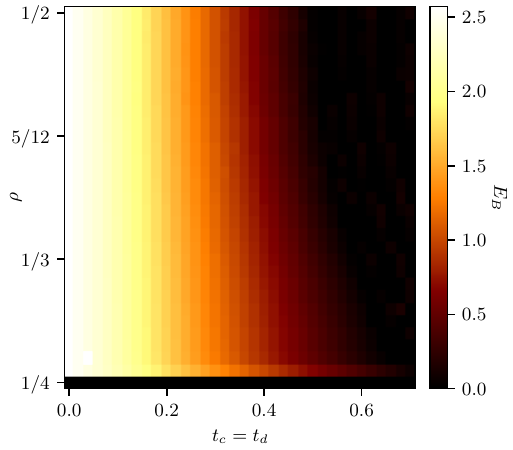}
    \caption{Pair binding energy \cw{in the large Hund's regime} as a function of fillling $\rho$ and intralayer hoppings $t_d = t_c$. The other parameters are fixed to $J=3$, $U_c=0$, $U_d=1$, $U^\prime=0.4$ and $t_\perp$ is chosen to maximize the local energy gain \cw{($t_\perp = \sqrt{J}/2 \cdot \sqrt{4J + |U_d - 2U^\prime|}$)}
    . The pair binding energy decreases with larger intralayer hoppings and larger fillings. For $t_c=t_d\rightarrow 0$, the pair binding energy approaches twice the local energy gain of a single rung. Superconductivity can only emerge for $t_c=t_d>0$, since the rungs are not correlated for $t_c=t_d=0$.}
    \label{fig:smJ}
\end{figure}

In this appendix we study the pair binding energy as a function of filling, for large Hund's couplings $J$. In Fig.~\ref{fig:smJ} we show the pair binding energy in dependence on $t_d=t_c$ and the density $\rho$. The pair binding energy is positive only for $\rho>1/4$, since for $\rho \leq 1/4$ only the magnetic layer is occupied and there are no bound Cooper pairs as explained in the main text. For fillings larger than $1/4$ the pair binding energy is nearly independent of the filling. This is expected for small intralayer hoppings t, since the pair binding energy approaches twice the local energy gain of a single rung for the one electron/hole states in the limit $t_c=t_d\rightarrow 0$. Interestingly, the behavior of the pair binding energy is very similar to the results in Fig. 4(a) of the main text, for small $J$.

\section{Superconducting Phase Diagrams and Form of Pair Correlation Function}
\label{app:sc_phase_diag}

\begin{figure}[t]
    \centering
    \includegraphics[width=1.0\columnwidth]{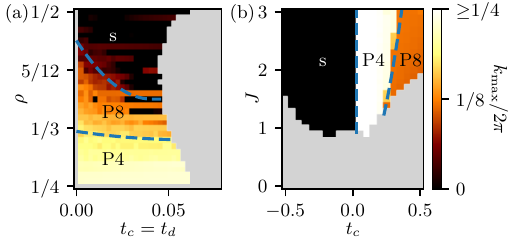}
    \caption{Superconducting phase diagram extracted via a Fourier transform around $R=0$ as defined for (a) $J=0.45$, $U^\prime=0.5$, $U_d=1.0$, $U_c=0.0$ and $t_\perp$ chosen to maximize the local energy gain as a function of the interlayer hoppings $t_c=t_d$ and the filling (\cw{small Hund's regime,} corresponding to Fig.~4 (a) of the main text);
    and for (b) $\rho=3/8$, $t_d=0.25$, and $U^\prime=0.4$ (other parameters as in (a)) as a function of $t_c$ and $J$ (corresponding to Fig.~4 (b) of the main text).
    In both cases we used the correlation functions for system size $L=60$ to measure the periodicity. The blue dashed lines show approximate divisions between an $s$-wave and a pair density wave (FFLO superconductor) with a periodicity of 4 or 8 rungs. While there is a continuous change of the extracted pairing momentum $k_\mathrm{max}$ in (a), we find that $k_\mathrm{max}$ jumps in (b) which leads to a discrete set of FFLO phases. Bond dimensions between 400 and 3200 were used, depending on the parameters.}
    \label{fig:sc_phases}
\end{figure}

In this appendix we show superconducting phase diagrams for additional parameter sets and discuss the phases indicated in Figs.~4 (a), (b) of the main text. We use a Fourier transform of the pair correlation function with $L=60$ modified with a gaussian window function $g(R) = \exp(-R^2/\sigma^2)$ with $\sigma=20$ around the center $R=0$ of the correlation function We further pad with zeros to smoothen the final result; the correlation function is already decayed to nearly zero at long distances due to the window function. We then plot the momentum with maximal value of the Fourier transform and define this value $k_\mathrm{max}$.
In Figs.~\ref{fig:sc_phases} (a) and (b) we show $k_\mathrm{max} / 2\pi$ for Figs.~4 (a) and (b) of the main text. In grey we display the regions where no superconductivity occurs. Here, we define a parameter set as superconducting if the pair binding energy is above a certain threshold ($E_B>0.003$ for (a) and $E_B>0.05$ for (b)). The threshold is used to avoid including points which have a finite pair binding energy due to finite size effects or numerical uncertainties. We further grey out the parameters with $t_c=t_d=0$. Although the pair binding energy is finite here, this does not lead to superconductivity since the different rungs are not coupled. 
The phase diagrams in Figs.~\ref{fig:sc_phases} are only qualitatively converged with the bond dimensions. Individual points with poor convergence are expected to be unimportant for this qualitative analysis. 

In Fig.~\ref{fig:sc_phases} (a) we find that the value of $k_\mathrm{max}$ changes continuously as a function of filling, suggesting incommensurate FFLO order. We note that for intermediate intralayer hoppings and intermediate filling, the value of $k_\mathrm{max}$ is hard to converge due to finite system size. Therefore, a unique definition of the FFLO periodicity is challenging and we only separate regions with approximate 4-fold and 8-fold periodicity via dashed blue lines, as guides to the eye.
In stark contrast we find discrete phases in Fig.~\ref{fig:sc_phases} (b) at fixed filling. Here, it is possible to uniquely identify the regimes with FFLO superconducting ground states with sharp differences in periodicity.

\begin{figure}[t]
    \centering
    \includegraphics[width=1.0\columnwidth]{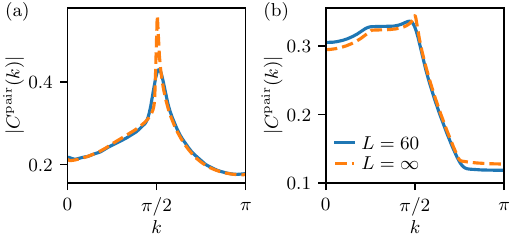}
    \caption{Fourier transform of the pair correlation function for $\rho = 3/8$, $t_d = 0.25$, $J = 3$, ${\color{black}U^\prime} = 0.4$, $U_d = 1$, $Uc = 0$, $t_\perp$ chosen to maximize the local energy gain and (a) $t_c=0.05$,  (b) $t_c=0.25$. While in (a) there is a very clear peak at $k=\pi/2$, the correlation function in (b) is broadened with a maximum at $k=\pi/2$.}
    \label{fig:corr_peaks}
\end{figure}

We note that the peaks of the Fourier transform of the pair correlation are not always well pronounced. In Fig.~\ref{fig:corr_peaks} we show the results for two exemplary values in the $k=\pi/2$ pair density wave phase. We see that for some parameter sets the peak is very dominant while for other parameter sets $k_\mathrm{max}$ is hard to identify. We also compare this with the results obtained with $L=\infty$ with iDMRG (for the infinite system we used a window function with $\sigma=100$ instead of $\sigma=20$) and see that the differences are rather small and that finite sizes effects cannot be the reason for lacking of a clear peak.

For these reasons, we employ a second method for the calculation of the momentum dependent pair correlation function. Instead of applying the Fourier transform around the center of the correlation at $R=0$, we calculate the Fourier transform in the tail of the correlation function. This method allow us to calculate directly from the long range behavior of the correlator without worrying about any short range additives or a different period around $R=0$. 
Additionally, we will get rid of the power-law decay, which ensures that the envelope does not falsify the results. 
The momentum dependent correlation function from the tail is calculated as follows: 
We fit a power-law to the real space correlation function and multiply the correlator with the power-law in order to have an approximately not decaying function. We then apply a gaussian window function in the middle of this extracted tail. Here, we use the correlation functions \cw{$C_{15, x}^{\rm pair}$} of a system with $L=60$ and center the window function at $x=30$ with a gaussian width of $\sigma=10$. Again, we can further add zeros at sides to smoothen the final result. Finally, we apply a Fourier transform to convert our results into momentum space. However, a strong disadvantage is that it is significantly harder to converge the results obtained with this method, since the tail of the powerlaw-decaying correlation function is less well-converged than its short-range behavior, at equal bond dimensions.

\begin{figure}[t]
    \centering
    \includegraphics[width=1.0\columnwidth]{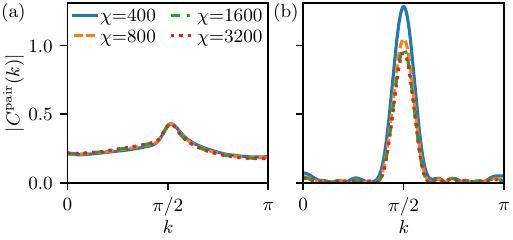}
    \caption{Fourier transform of the pair correlation function for $\rho = 3/8$, $t_c=0.05$, $t_d = 0.25$, $J = 3$, ${\color{black}U^\prime} = 0.4$, $U_d = 1$, $U_c = 0$ and $t_\perp$ chosen to maximize the local energy gain with $L=60$ and multiple bond dimensions. (a) The Fourier transform is performed around $R=0$. The peak at $\pi/2$ indicates a periodicity of 4 rungs. 
    (b) Fourier transform of the tail of the correlation function, after dividing the correlation function with its power-law decay, recovering the same peak as in (a).}
\label{fig:both_work}
\end{figure}

{\color{black}Although the results from both methods can strongly disagree as shown below, they coincide across a significant part of the phase diagram, leading to a similar momentum-dependent pair correlation function. In Fig.~\ref{fig:both_work} we show the results for a parameter set, where the results do agree. Additionally, for this parameter set both peaks are well converged with bond dimension. Nevertheless, the results from the tail are not always that well converged due to the reasons mentioned above.}

\begin{figure}[t]
    \centering
    \includegraphics[width=1.0\columnwidth]{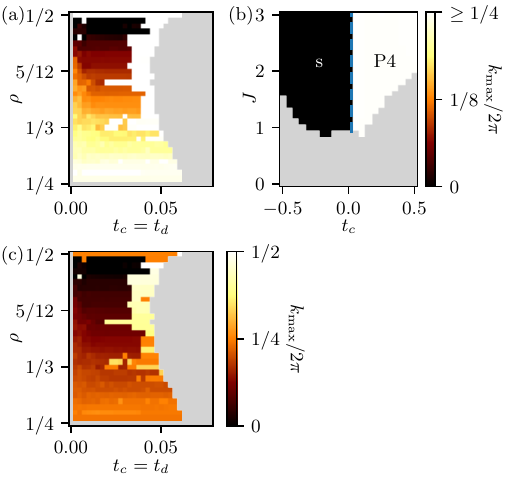}
    \caption{Superconducting phase diagram extracted via a Fourier transform of the tail of the pair correlation function for the same parameters as in Fig.~\ref{fig:sc_phases}, for a system size of $L=60$. (a) In agreement with Fig.~\ref{fig:sc_phases}(a) \cw{(small Hund's regime)}, the momentum of the peak changes continuously upon changing the filling fraction. The pairing momentum agrees with the analysis of the correlation function around $R=0$ in Fig.~\ref{fig:sc_phases} for small hopping amplitudes (the strong coupling regime) but becomes pinned to $k\geq\pi/2$ for large hoppings. (c) Same results with a different colorbar.
    (b) Results for the same parameters as Fig.~\ref{fig:sc_phases}(b). We still find an phase transition between the standard s-wave superconductivity and an FFLO phase. In contrast to Fig.~\ref{fig:sc_phases}(b), we find a periodicity of 4 rungs for the whole FFLO regime.
    Bond dimensions between 400 and 3200 were used, depending on the parameters.}
    \label{fig:sc_phases_tail}
\end{figure}

Let us now investigate the influence of the different methods for the calculation of the superconducting phase diagrams. We calculate the phase diagrams found in Fig.~\ref{fig:sc_phases} again using the tail of the correlation functions and show the results in Fig.~\ref{fig:sc_phases_tail}. 
Let us first investigate the differences of Fig.~\ref{fig:sc_phases}(a) and Fig.~\ref{fig:sc_phases_tail}(a) as well as Fig.~\ref{fig:sc_phases_tail}(c), where the results are shown with a different colorbar. Here, we note a similar behavior for small $t_c=t_d$. We still find a finite pairing momentum which is slowly changing from $k_{\rm max}=\pi$ to $0$ upon increasing the filling of the system. For larger $t_c=t_d$ we instead find a large and stable regime with a $2$-rung periodic FFLO phase.
In (b), there is again a significant region, where the phases agree for both methods. Only the P8 phase for large $t_c>0$ disappears and the P4 phase gets enlarged. For these parameters, we find a conventional s-wave for $t_c\leq 0$ and a 4-rung periodic FFLO phase for $t_c>0$. The phase diagram seems to be qualitatively converged, but we stress again that, there could still be changes in the maximum for even higher bond dimensions \cw{due to the poor convergence of the tail.}
For both panels, we note that the results can differ in comparison to the results shown in Fig.~\ref{fig:sc_phases} with a Fourier transform around $R=0$. Therefore, we are not able to assign a clear pairing momentum to all parameter sets. For (b), we can still tell that a conventional $s$-wave appears for $t_c\leq 0$, while we find an FFLO phase for $t_c>0$ as we have marked in Fig.~4(b) of the main text. 

Finally, let us quickly comment how the pairing momentum phase diagrams relate to the alternate method of verification of superconducting order using small artificial pairing fields, as described in the main text. The periodicity of the seeded order parameter should be chosen according to the periodicity of the pair correlation function. For the parameters used for the calculations with the artificial pairing field in the main text (along the cut $t_c=0.25$ of Figs.~\ref{fig:sc_phases}(b) and \ref{fig:sc_phases_tail}(b)), the pairing momentum phase diagrams Fig.~\ref{fig:sc_phases}(b) and Fig.~\ref{fig:sc_phases_tail}(b) agree show a FFLO phase with a periodicity of 4 rungs. For small Hund's coupling, the result in Fig.~\ref{fig:sc_phases}(b) differs and suggests a 8-rung periodic PDW, however, the peaks of the Fourier transform (applied around $R=0$) of the pair correlation function in this region are very broad with similar weight at $k=\pi/2$ and $k=\pi/4$ belonging to a 4-rung and 8-rung periodic PDW, respectively.
Assuming a 4-rung periodic FFLO phase for the definition of an artificial pairing field is therefore consistent with both presented evaluations of pair correlation function.
The validity of this assumption can also be seen by investigating the results in Fig.~4(d) of the main text. Here, we note that \cw{the results extracted from} the order parameter found with a small artificial pairing field perfectly agree with the point where the pair binding energy becomes finite \cw{and where the pairing strength shows a jump}.

\section{Behavior of order parameter with applied bias field in the limit $h\rightarrow 0$}
\begin{figure}[t]
    \centering
    \includegraphics[width=1.0\columnwidth]{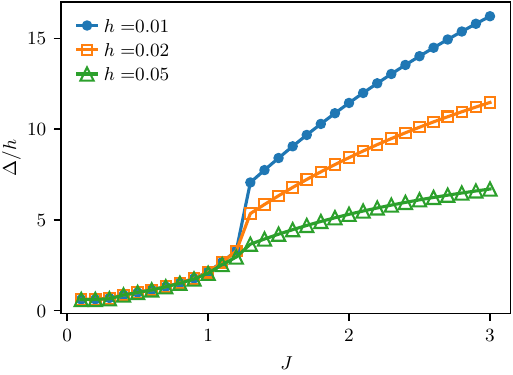}
    \caption{$\Delta/h$ for different pairing field strengths $h$ as a function of the Hund's exchange $J$. The other parameters are $t_c=t_d=0.25$, $U_c=0$, $U_d=1$, $U^\prime=0.4$ and $t_\perp$ chosen to maximize the local energy gain. The different curves lie on top of each other for $J\lesssim1.2$. Then, they show a jump that gets steeper for smaller $h$. This provides further evidence that the system undergoes a superconducting phase transition upon increasing $J$.}
    \label{fig:jump_steep}
\end{figure}

In this appendix we discuss how the superconducting order parameter $\Delta$ behaves for different pairing field strengths $h$. In Fig.~4(d) of the main text the superconducting phase transition results in a jump of the order parameter for fixed $h$. We now renormalize the order parameter by the pairing field strength and compare it for different $h$. The parameters are chosen corresponding to Fig.~4\cw{(d)} of the main text. We show the results in Fig.~\ref{fig:jump_steep} and see that the jump in $\Delta / h$ increases for smaller $h$. In the limit of a disappearing pairing field this leads to a diverging jump what is a strong evidence for a phase transition at this point, to a 1D superconducting state with quasi-long-range order.

\begin{figure}[t!]
    \centering
    \includegraphics[width=1.0\columnwidth]{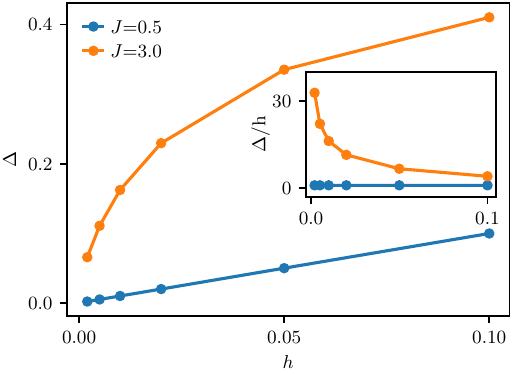}
    \caption{Scaling behavior of the order parameter $\Delta$ as a function of the pairing field strength $h$ for $J=0.5$ (deep in the {\color{black}normal} phase) and $J=3.0$ (deep in the superconducting phase). The order parameter $\Delta$ becomes considerably larger in the superconducting regime, and importantly exhibits a jump at the phase transition point found in Fig.~4(d) of the main text and Fig.~\ref{fig:jump_steep} of the supplement. Furthermore, the scaling of the two curves differs. For $J=0.5$, the order parameter increases linearly with $h$, while the results for $J=3.0$ follow a power-law with an exponent smaller than one. In the inset we show the scaling for $\Delta/h$, showing that $\Delta/h$ diverges for $h\rightarrow 0$.
    \cw{The identification of superconducting regimes with quasi-long-range order using pair field bias calculations are broadly consistent with the additional described metrics (pair binding energy and pairing strength based on the pairing correlation functions). }
    }
    \label{fig:delta_scaling}
\end{figure}

In Fig.~\ref{fig:delta_scaling} the order parameter is shown as a function of the pairing field strength $h$, both for a parameter point which we previously identified (via the pair binding energy and pair correlation functions) to host a superconducting phase, and for a parameter point without superconductivity. The response to the pairing field is significantly enhanced in the superconducting regime; more importantly, we can see that the two curves scale differently with $h$. While $\Delta$ scales linearly with $h$ in the normal phase, the superconducting regime shows that $\Delta$ follows a power law with an exponent smaller than one for small pairing field strengths, leading to a divergence of $\lim_{h\rightarrow 0}\Delta/h$ as shown in the inset. We note that the $\Delta$ must approach zero for a vanishing pairing field due to the Mermin-Wagner theorem.

\section{Form of Magnetic Correlation Function}
In this section, we discuss how the magnetic phase diagrams in Figs.~3(a),(b),(c) were determined and how the different phases are defined. First, we applied a Fourier transform around $R=0$, analogous to our analysis of the pair correlation function in Appendix~\ref{app:sc_phase_diag}. In contrast to the superconducting phases, iDMRG calculations were used to find the spin correlation function in the thermodynamic limit {\color{black}and a gaussian window with $\sigma=100$ was applied}. 
We then assign a $2$-{\color{black}rung} or $4$-{\color{black}rung} AFM phase if we find the maximum of the correlation function at $k=\pi$ or $k=\pi/2$, respectively. We also find a phase, where there is no clear peak in the momentum dependent correlation, which we denote as disordered. We define a phase as disordered if the maximal \cw{peak} value is less than 1.5 times (small changes to this value do not change the phase diagrams significantly) as large as the value at $k=\pi$ \cw{for the same parameter set}.
Later on, we will show an example of these correlations making this choice understandable. 
We furthermore also identify points in the phase diagram as magnetically disordered if their correlation function does not decay with a power-law but with an exponential decay. 
In order to decide whether a decay follows a power-law or an exponential decay, we fit both an exponential and a powerlaw ansatz \cw{to the extrapolated data at infinite bond dimension} and compare the squared deviation with respect the correlation function tail.
The squared deviation is summed over even distances $R$ of the correlation function since we are only interested in the envelope function and want to avoid additional variations due to a possible 4-rung periodicity (2-rung periodicity of the absolute value).
\cw{The range for the fit is determined via a maximal deviation of the result in an logarithmic scaling with base 10  between the the smallest and largest bond dimensions which were used for the extrapolation.}
In Fig. 3(c) of the main text (the magnetic phase diagram for half filling), this leads to large regions of magnetic disorder. 
We compared results found with $\chi=200, 400, 800$ or $\chi=400, 800, 1600$ \cw{and a range defined using a deviation of 1 or 2.}
in order to make sure that the phase diagram is converged. Parameters around the phase boundary, where the results do not agree \cw{or where no clear phase could be assigned to the numerical data}, are greyed out.

For quarter filling, the comparison of different ranges and bond dimensions sets is not necessary, since an exponential decay only occurs at phase transition points, where simulations become poorly converged, and deep within the phase already identified as disordered due to their momentum dependent correlation function. Here, we only used $\chi=200, 400, 800$ and \cw{determined the used range via an allowed deviation of 2.}
We mark regions as grey, where it is not possible to assign a clear phase from numerical results. Furthermore, we grey out the regime $t_d=t_c=0$ since in this case there are no correlations between the rungs by construction.

Finally, we find a phase separation regime shown in Fig.~3(b) of the main text. To ascertain phase separation, we complementarily used a chain with 8 rungs and periodic boundary conditions and checked the energy difference of the lowest phases. The parameter sets are assigned to the phase separation regime if the energy difference between states with different ordering periodicities was quasi-degenerate (using a threshold of $10^{-4.5} \cw{U_d}$ due to finite size effects).
More details on this can be found in Appendix \ref{app:phase_sep}.

\begin{figure}[t!]
    \centering
    \includegraphics[width=1.0\columnwidth]{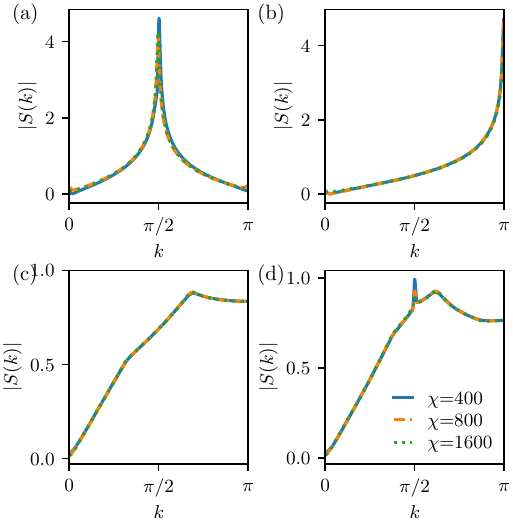}
    \caption{Magnetic correlations at quarter filling with $t_d, t_c, J =$ (a) $0.25, -0.25, 1.5$; (b) $0.25, 0.25, 1.5$; (c) $0.5, 0.5, 0.1$; and (d) $0.5, -0.5, 0.3$ with multiple bond dimensions  showing exemplary values of the different phases (a) 4-{\color{black}rung} AFM, (b) 2-{\color{black}rung} AFM, (c) disordered. In (d) it is not possible to unambiguously assign a phase.}
    \label{fig:mag_corr_quarter}
\end{figure}

\begin{figure}[t!]
    \centering
    \includegraphics[width=1.0\columnwidth]{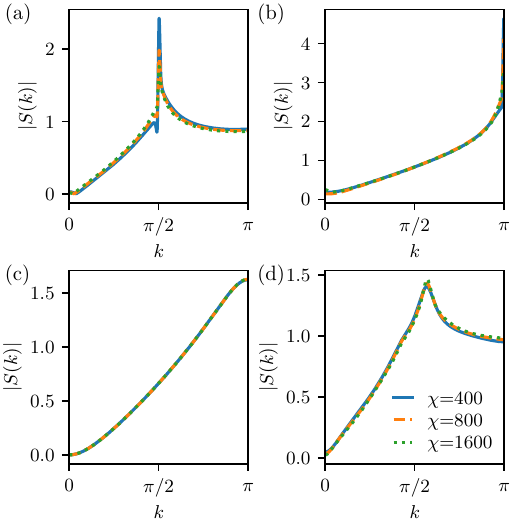}
    \caption{Magnetic correlations at half filling with $t_d, t_c, J =$ (a) $0.25, 0.25, 2.0$; (b) $0.25, -0.5, 0.0$; (c) $0.25, -0.25, 1.5$ and (d) $0.25, 0.25, 1.0$ with multiple bond dimensions showing exemplary values of the different phases (a) 4-{\color{black}rung} AFM, (b) 2-{\color{black}rung} AFM, (c),(d) disordered.}
    \label{fig:mag_corr_half}
\end{figure}

In Figs.~\ref{fig:mag_corr_quarter} and \ref{fig:mag_corr_half} we show the momentum dependent magnetic correlation functions for exemplary values of the different phases at quarter and half filling, respectively. In Fig.~\ref{fig:mag_corr_quarter}(a) and (b), we see the Fourier transform of the pair correlation function of parameters in the 4-rung AFM and 2-rung AFM phase with clear and converged peaks at the corresponding momentum points. In (c), we show the results within the paramagnetic phase, where there is no clear peak visible. Parameter sets with correlation functions as shown in (d) are not assigned a phase, as the small peak at $k=\pi/2$ decreases with larger bond dimensions. In Fig.~\ref{fig:mag_corr_half}(a) we show an exemplary parameter set from the 4-{\color{black}rung} AFM phase with a peak at $k=\pi/2$. In (b), we show the Fourier transform for the 2-{\color{black}rung} AFM phase with a very pronounced and converged peak at $k=\pi$. In (c) and (d) no clear peaks are visible. Here, the magnetic correlations also decay exponentially, assigning both parameter sets to the disordered phase. We also applied the same analysis to a Fourier transform of the tail of the magnetic correlation functions (in analogy to our analysis of pairing correlations). The phase diagram takes the same form, and we therefore refrain from showing these results here.

\section{Phase Separation in the magnetic phase diagram}
\label{app:phase_sep}

\begin{figure}[t!]
    \centering
    \includegraphics[width=1.0\columnwidth]{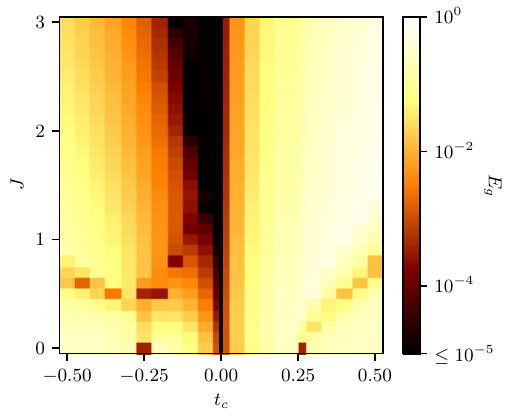}
    \caption{False color plot showing $E_g$ as a function of $t_c$ and $J$. $E_g$ is given by the difference between the ground state and the first excited state with a different magnetic behavior, computed in exact diagonalization for 8 rungs with periodic boundary conditions. The other parameters are given by $L=8$, $\rho=1/4$, $t_d=|t_c|$, $J=3.0$, $V=0.4$, $U_d=1.0$, $U_c=0.0$ and $t_\perp$ chosen to maximize the local energy gain via charge disproportion (corresponding to Fig.~3(b) of the main text). For small $t_c=-t_d$ the different states are quasi-degenerate what supports our conclusion of phase separation in this regime. We also see that $E_g$ shows minima at lines which corresponds to the phase transitions shown in Fig.~3(b) of the main text up to finite size effects.}
    \label{fig:ED_gap}
\end{figure}

In this appendix, we provide additional detail on identifying the phase separation regime found in Fig.~3(b) of main text. When doing (i)DMRG calculations one finds them to fail to properly describe the ground state inside this regime.
The \cw{periodicity of the correlation functions} crucially depend on the exact parameters (e.g. bond dimension or system size).
\cw{This indicates that DMRG accidentally falls into one of multiple (nearly) degenerate ground states with different magnetic ordering.} 
Therefore, we are not able to define the magnetic phase in that region. This failure of DMRG is already a first evidence for a phase separation.
To illustrate the tendency for phase separation, we provide exact diagonalization (ED) results for a periodic chain with 8 rungs.
In Fig.~\ref{fig:ED_gap} we show the energy difference $E_g$ of the ground state with the first excited state with a different magnetic ordering.
\cw{Here, we defined the phase of every state by comparing the values of the Fourier transform at $k=0$ (FM), $k=\pi$ (2-{\color{black}rung} AFM) and $k=\pi/2$ (4-{\color{black}rung} AFM) and assigning the phase corresponding to the maximal value.}
Indeed, this energy difference is extremely small ($\leq 10^{-4.5} \cw{U_d}$)
in the regime in question. Furthermore, we were able to see that $E_g$ decreases with increasing system size. This leads to the conclusion that different magnetic phases are indeed degenerate for small $t_c=-t_d$ and large $J$ resulting in phase separation. Additionally, we want to stress that other minima of $E_g$ are visible in Fig.~\ref{fig:ED_gap}. These belong to the phase transitions found in Fig.~3(b) in the main text and show that our analysis is consistent. Small differences of the phase boundaries occur due to finite size effects. 

\section{Strong Coupling Theory of Intertwined Magnetism and Superconductivity}

The main text describes the derivation of an effective strong coupling theory of $S=1/2$ magnetic moments coupled to charge-$2e$ Cooper pairs, taking the form of an effective Kugel-Khomskii spin-orbital model upon replacing charge-$2e$ hardcore bosons by $S=1/2$ pseudospins. The Monte Carlo simulations presented in the main text study emergent phases of this model. Here, we provide the perturbative expressions (to second order in intralayer hopping $t_c$, $t_d$) of the strong coupling parameters. In full generality, the strong-coupling expansion describes the interaction of spin $S=1/2$ moments \textit{and} charge-$2e$ hardcore bosons, both of which are delocalized across rungs of the bilayer, making the rigorous perturbative strong-coupling analysis a lengthy endeavor. However, we note that, for $U_d \gg U_c$, the $S=1/2$ magnetic moment is hosted primarily in the strongly-interacting layer of ``d'' fermions, whereas the charge-$2e$ Cooper pairs live primarily in the metallic layer of ``c'' fermions. Below, we quote the perturbative expressions without approximation, which describe the effective spin-pairing model parameters in all generality as a function of $t_c, t_d, U_c, U_d, U', J$ and $t_\perp$. 
\begin{widetext}
We define
\begin{align}
    U &= \frac{U_d + U_c}{2} \\
    t &= t_d - t_c \\
    T &= t_d + t_c \\
    r &= \frac{\sqrt{[16 t_\perp^2 + (U_d - U_c)^2] [ 64 t_\perp^2 + (U_d+U_c-2U')^2]}}{32 t_\perp} \\
    \tan(\phi) &= \frac{8 t_\perp}{U_d + U_c - 2U'} \\
    \tan(\theta) &= \frac{4 t_\perp}{U_d - U_c} \\
    \epsilon &= \frac{U_d + U_c}{2} + J + 2 r \sin(\phi)
\end{align}
and obtain
\begin{align}
    J_{s} &= -\frac{\left(\sin \left(\frac{\phi }{2}\right) (t \cos (\theta )-T)+T \sin (\theta ) \cos \left(\frac{\phi }{2}\right)\right)^2}{4 (-2 (J+U)+2 r \sin (\theta ) (\cos (\phi )+1)+\epsilon )}-\frac{\left(\cos \left(\frac{\phi }{2}\right) (T-t \cos (\theta ))+T \sin (\theta ) \sin \left(\frac{\phi
   }{2}\right)\right)^2}{4 (-2 (J+U)+2 r \sin (\theta ) (\cos (\phi )-1)+\epsilon )} \notag\\
   &-\frac{(t (-\sin (\theta )) \sin (\phi )+t+T \cos (\theta ) \cos (\phi ))^2}{16 \left(-4 J+8 r \sin (\theta ) \sin ^2\left(\frac{\phi }{2}\right)+\epsilon \right)}+\frac{\left(\cos \left(\frac{\phi }{2}\right) (t \cos (\theta
   )+T)-T \sin (\theta ) \sin \left(\frac{\phi }{2}\right)\right)^2}{8 \left(-2 J+4 r \sin (\theta ) \sin ^2\left(\frac{\phi }{2}\right)+\epsilon \right)} \notag\\
   &-\frac{\left(\sin \left(\frac{\phi }{2}\right) (t \cos (\theta )-T)+T \sin (\theta ) \cos \left(\frac{\phi }{2}\right)\right)^2}{8 (-2 J+2 r \sin (\theta )
   (1-3 \cos (\phi ))+\epsilon )}+\frac{\left(\sin \left(\frac{\phi }{2}\right) (t \cos (\theta )+T)+T \sin (\theta ) \cos \left(\frac{\phi }{2}\right)\right)^2}{8 (-2 J-2 r \sin (\theta ) (\cos (\phi )+1)+\epsilon )} \notag\\
   &-\frac{\left(\cos \left(\frac{\phi }{2}\right) (T-t \cos (\theta ))+T \sin (\theta ) \sin
   \left(\frac{\phi }{2}\right)\right)^2}{8 (-2 J-2 r \sin (\theta ) (3 \cos (\phi )+1)+\epsilon )}-\frac{(t \sin (\theta ) \cos (\phi )+T \cos (\theta ) \sin (\phi ))^2}{8 (-4 J-4 r \sin (\theta ) \cos (\phi )+\epsilon )} \notag\\
   &-\frac{(t \sin (\theta ) \sin (\phi )+t-T \cos (\theta ) \cos (\phi ))^2}{16 (-4 J-4 r \sin
   (\theta ) (\cos (\phi )+1)+\epsilon )}+\frac{t^2 \sin ^2(\theta )}{4 (4 r \sin (\theta ) \cos (\phi )-2 U+\epsilon )} +\frac{t^2 \sin ^2(\theta )}{8 (\epsilon -4 r \sin (\theta ) \cos (\phi ))} \notag\\
   &-\frac{(t-T \cos (\theta ))^2}{4 (8 r \sin (\theta ) \cos (\phi )-4 U+\epsilon )}-\frac{(t-T \cos (\theta ))^2}{16
   (\epsilon -8 r \sin (\theta ) \cos (\phi ))}-\frac{(t-T \cos (\theta ))^2}{4 (\epsilon -2 U)}-\frac{(t+T \cos (\theta ))^2}{16 \epsilon } \\
    K_{xy} &= \frac{(t (-\sin (\theta )) \sin (\phi )+t+T \cos (\theta ) \cos (\phi ))^2}{16 \left(-4 J+8 r \sin (\theta ) \sin ^2\left(\frac{\phi }{2}\right)+\epsilon \right)}+\frac{3 \left(\cos \left(\frac{\phi }{2}\right) (t \cos (\theta )+T)-T \sin (\theta ) \sin \left(\frac{\phi }{2}\right)\right)^2}{8 \left(-2 J+4 r
   \sin (\theta ) \sin ^2\left(\frac{\phi }{2}\right)+\epsilon \right)} \notag\\
   &-\frac{\left(\sin \left(\frac{\phi }{2}\right) (t \cos (\theta )-T)+T \sin (\theta ) \cos \left(\frac{\phi }{2}\right)\right)^2}{8 (-2 J+2 r \sin (\theta ) (1-3 \cos (\phi ))+\epsilon )}+\frac{3 \left(\sin \left(\frac{\phi }{2}\right) (t \cos
   (\theta )+T)+T \sin (\theta ) \cos \left(\frac{\phi }{2}\right)\right)^2}{8 (-2 J-2 r \sin (\theta ) (\cos (\phi )+1)+\epsilon )} \notag\\
   &-\frac{\left(\cos \left(\frac{\phi }{2}\right) (T-t \cos (\theta ))+T \sin (\theta ) \sin \left(\frac{\phi }{2}\right)\right)^2}{8 (-2 J-2 r \sin (\theta ) (3 \cos (\phi
   )+1)+\epsilon )}+\frac{(t \sin (\theta ) \cos (\phi )+T \cos (\theta ) \sin (\phi ))^2}{8 (-4 J-4 r \sin (\theta ) \cos (\phi )+\epsilon )} \notag\\
   &+\frac{(t \sin (\theta ) \sin (\phi )+t-T \cos (\theta ) \cos (\phi ))^2}{16 (-4 J-4 r \sin (\theta ) (\cos (\phi )+1)+\epsilon )}-\frac{3 t^2 \sin ^2(\theta )}{8
   (\epsilon -4 r \sin (\theta ) \cos (\phi ))} +\frac{(t-T \cos (\theta ))^2}{16 (\epsilon -8 r \sin (\theta ) \cos (\phi ))}+\frac{9 (t+T \cos (\theta ))^2}{16 \epsilon } \\
    K_{z} &= J_s + \frac{\left(\sin \left(\frac{\phi }{2}\right) (t \cos (\theta )-T)+T \sin (\theta ) \cos \left(\frac{\phi }{2}\right)\right)^2}{2 (-2 (J+U)+2 r \sin (\theta ) (\cos (\phi )+1)+\epsilon )}+\frac{\left(\cos \left(\frac{\phi }{2}\right) (T-t \cos (\theta ))+T \sin (\theta ) \sin \left(\frac{\phi
   }{2}\right)\right)^2}{2 (-2 (J+U)+2 r \sin (\theta ) (\cos (\phi )-1)+\epsilon )} \notag\\
   &-\frac{\left(\cos \left(\frac{\phi }{2}\right) (t \cos (\theta )+T)-T \sin (\theta ) \sin \left(\frac{\phi }{2}\right)\right)^2}{2 \left(-2 J+4 r \sin (\theta ) \sin ^2\left(\frac{\phi }{2}\right)+\epsilon
   \right)}-\frac{\left(\sin \left(\frac{\phi }{2}\right) (t \cos (\theta )+T)+T \sin (\theta ) \cos \left(\frac{\phi }{2}\right)\right)^2}{2 (-2 J-2 r \sin (\theta ) (\cos (\phi )+1)+\epsilon )} \notag\\
   &+\frac{t^2 \sin ^2(\theta )}{2 (4 r \sin (\theta ) \cos (\phi )-2 U+\epsilon )}-\frac{t^2 \sin ^2(\theta )}{2
   (\epsilon -4 r \sin (\theta ) \cos (\phi ))}+\frac{(t-T \cos (\theta ))^2}{2 (\epsilon -2 U)}-\frac{(t+T \cos (\theta ))^2}{2 \epsilon } \\
    W_{xy} &= J_s + \frac{\left(\sin \left(\frac{\phi }{2}\right) (t \cos (\theta )-T)+T \sin (\theta ) \cos \left(\frac{\phi }{2}\right)\right)^2}{4 (-2 (J+U)+2 r \sin (\theta ) (\cos (\phi )+1)+\epsilon )}+\frac{\left(\cos \left(\frac{\phi }{2}\right) (T-t \cos (\theta ))+T \sin (\theta ) \sin \left(\frac{\phi
   }{2}\right)\right)^2}{4 (-2 (J+U)+2 r \sin (\theta ) (\cos (\phi )-1)+\epsilon )} \notag\\
   &+\frac{\left(\sin \left(\frac{\phi }{2}\right) (t \cos (\theta )-T)+T \sin (\theta ) \cos \left(\frac{\phi }{2}\right)\right)^2}{4 (-2 J+2 r \sin (\theta ) (1-3 \cos (\phi ))+\epsilon )}+\frac{\left(\cos \left(\frac{\phi
   }{2}\right) (T-t \cos (\theta ))+T \sin (\theta ) \sin \left(\frac{\phi }{2}\right)\right)^2}{4 (-2 J-2 r \sin (\theta ) (3 \cos (\phi )+1)+\epsilon )} \notag\\
   &-\frac{t^2 \sin ^2(\theta )}{4 (4 r \sin (\theta ) \cos (\phi )-2 U+\epsilon )}-\frac{t^2 \sin ^2(\theta )}{4 (\epsilon -4 r \sin (\theta ) \cos (\phi
   ))}+\frac{(t-T \cos (\theta ))^2}{4 (8 r \sin (\theta ) \cos (\phi )-4 U+\epsilon )}+\frac{(t-T \cos (\theta ))^2}{4 (\epsilon -2 U)} \\
    W_{z} &= -\frac{\left(\sin \left(\frac{\phi }{2}\right) (t \cos (\theta )-T)+T \sin (\theta ) \cos \left(\frac{\phi }{2}\right)\right)^2}{4 (-2 (J+U)+2 r \sin (\theta ) (\cos (\phi )+1)+\epsilon )}-\frac{\left(\cos \left(\frac{\phi }{2}\right) (T-t \cos (\theta ))+T \sin (\theta ) \sin \left(\frac{\phi
   }{2}\right)\right)^2}{4 (-2 (J+U)+2 r \sin (\theta ) (\cos (\phi )-1)+\epsilon )}\notag\\
   &+\frac{(t (-\sin (\theta )) \sin (\phi )+t+T \cos (\theta ) \cos (\phi ))^2}{16 \left(-4 J+8 r \sin (\theta ) \sin ^2\left(\frac{\phi }{2}\right)+\epsilon \right)}-\frac{\left(\cos \left(\frac{\phi }{2}\right) (t \cos (\theta
   )+T)-T \sin (\theta ) \sin \left(\frac{\phi }{2}\right)\right)^2}{8 \left(-2 J+4 r \sin (\theta ) \sin ^2\left(\frac{\phi }{2}\right)+\epsilon \right)} \notag\\
   &+\frac{\left(\sin \left(\frac{\phi }{2}\right) (t \cos (\theta )-T)+T \sin (\theta ) \cos \left(\frac{\phi }{2}\right)\right)^2}{8 (-2 J+2 r \sin (\theta )
   (1-3 \cos (\phi ))+\epsilon )}-\frac{\left(\sin \left(\frac{\phi }{2}\right) (t \cos (\theta )+T)+T \sin (\theta ) \cos \left(\frac{\phi }{2}\right)\right)^2}{8 (-2 J-2 r \sin (\theta ) (\cos (\phi )+1)+\epsilon )} \notag\\
   &+\frac{\left(\cos \left(\frac{\phi }{2}\right) (T-t \cos (\theta ))+T \sin (\theta ) \sin
   \left(\frac{\phi }{2}\right)\right)^2}{8 (-2 J-2 r \sin (\theta ) (3 \cos (\phi )+1)+\epsilon )}+\frac{(t \sin (\theta ) \cos (\phi )+T \cos (\theta ) \sin (\phi ))^2}{8 (-4 J-4 r \sin (\theta ) \cos (\phi )+\epsilon )}\notag\\
   &+\frac{(t \sin (\theta ) \sin (\phi )+t-T \cos (\theta ) \cos (\phi ))^2}{16 (-4 J-4 r \sin
   (\theta ) (\cos (\phi )+1)+\epsilon )}+\frac{t^2 \sin ^2(\theta )}{4 (4 r \sin (\theta ) \cos (\phi )-2 U+\epsilon )}-\frac{t^2 \sin ^2(\theta )}{8 (\epsilon -4 r \sin (\theta ) \cos (\phi ))}\notag\\
   &+\frac{(t-T \cos (\theta ))^2}{4 (8 r \sin (\theta ) \cos (\phi )-4 U+\epsilon )}+\frac{(t-T \cos (\theta ))^2}{16
   (\epsilon -8 r \sin (\theta ) \cos (\phi ))}-\frac{(t-T \cos (\theta ))^2}{4 (\epsilon -2 U)}+\frac{(t+T \cos (\theta ))^2}{16 \epsilon }
\end{align}
\end{widetext}

\section{Details of the Monte Carlo simulations}

In this appendix, we explain in detail the Monte Carlo simulations, which were used for Fig.~5 and 6 of the main text.
We use the Monte Carlo method developed by some of us in Ref.~\cite{Gresista_2023} to study the strong-coupling Kugel-Khomskii Hamiltonian in two dimensions. Our simulations are classical in the sense that we assume the ground state of our model to be a product state $|\psi \rangle = \bigotimes_i |\psi_i \rangle$ where $i$ denotes a lattice site. To take into account the correct representation of the $\mathfrak{su}$(4) algebra for the system at hand, states $| \psi_i \rangle$ in the local Hilbert space must be characterized either by four (fundamental representation) or six (self-conjugate representation) complex numbers. This step has no classical analog, and we thus call our approach 'semi-classical'. Once a proper basis is chosen for the spin and pseudospin operators the classical energy can be minimized by standard means. Here, we use simulated annealing with Gaussian trial moves ($\sim 10^7$ sweeps) followed by stochastic gradient descent ($\sim 50$ sweeps) to obtain the ground state (see App.~A in Ref.~\cite{Gresista_2023} for details).

\section{Determination of the semi-classical phase diagram}

In this appendix, we discuss how the Monte-Carlo phase diagram presented in Fig.~6(b) of the main text was determined. Here, we analyzed spin and pseudospin magnetizations defined as

\begin{align}
    \langle \mathbf{S} \rangle &= \frac{1}{N} \sum_{i} \xi_i |(\langle S^x_i \rangle_C, \langle S^y_i \rangle_C, \langle S^z_i \rangle_C)^T| \notag \\ 
    \langle \tau^{\perp} \rangle &= \frac{1}{N} \sum_{i} \xi_i |(\langle \tau^x_i \rangle_C, \langle \tau^y_i \rangle_C)^T| \notag \\ 
    \langle \tau^z \rangle &= \frac{1}{N} \sum_{i} \xi_i |\langle \tau^z_i \rangle_C| \,,
    \label{eq:order_params}
\end{align}

where $\langle \mathcal{O}_i \rangle_C$ denotes the expectation value of an operator $\mathcal{O}$ at site $i$ for the optimized product wave function $C$. Here, $N$ is the total number of sites and $|.|$ denotes the euclidean norm. $\xi_i $ allows us to impose either a uniform ($\xi_i = \xi = \pm 1$) or staggered ($\xi = +1$ if $i$ 'even' \& $\xi = -1$ if $i$ 'odd') sign structure when measuring the site-averaged expectation value. This way, our calculations are sensitive to both ferro and antiferromagnetic correlations, respectively. Analyzing the corresponding two spin/pseudospin correlation functions, we confirmed that these expectation values are sufficient to characterize all phases in the parameter space we explore. 

\begin{figure*}
    \centering
    \includegraphics[width=\linewidth]{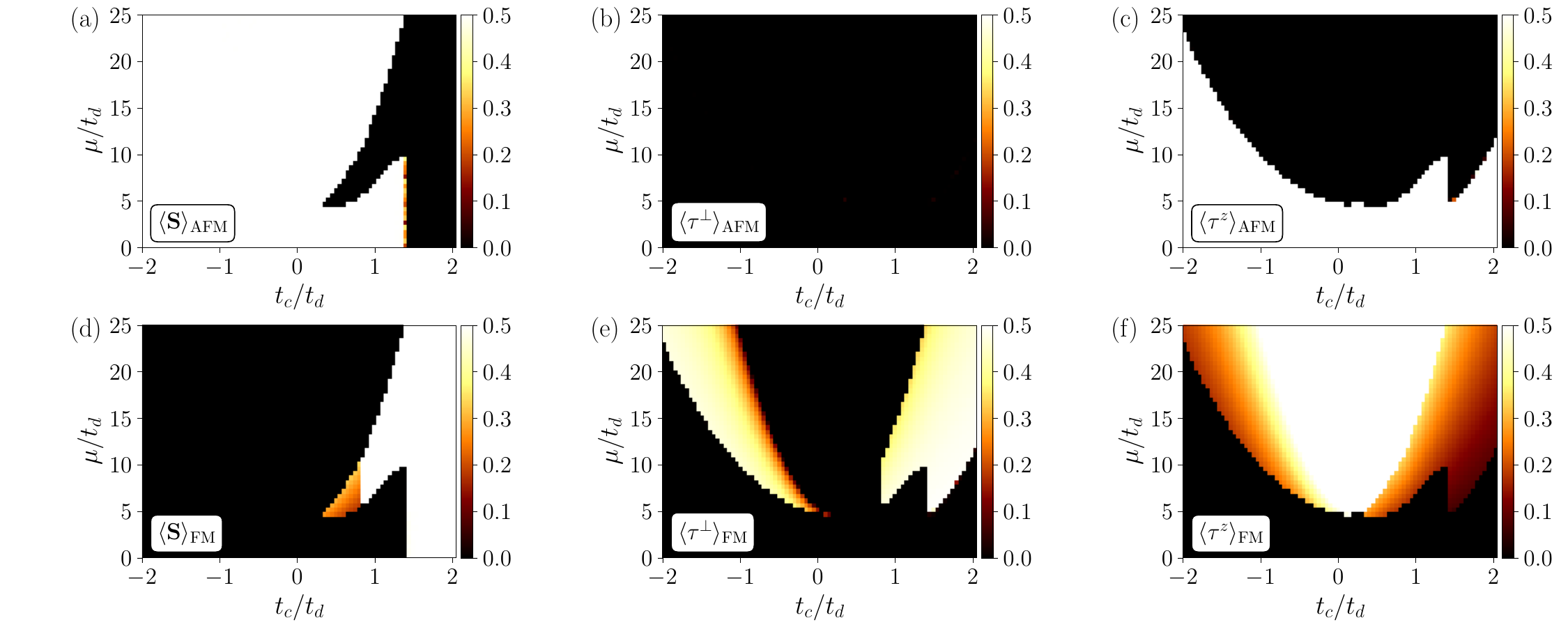}
    \caption{Measurement of the expectation values \eqref{eq:order_params} in semi-classical Monte Carlo. The data can be used to construct the phase diagram shown in Fig.~6(b) of the main text.}
    \label{fig:MC_magnetizations.pdf}
\end{figure*}

In Fig.~\ref{fig:MC_magnetizations.pdf} we show the obtained Monte Carlo data for the exchange couplings shown in Fig.~5 of the main text and at finite doping. Note that U(1) symmetry breaking, i.e.~a finite $\langle \tau^{\perp} \rangle$, occurs exclusively for finite field $\mu$ and for a partially-filled metallic layer ($\langle \tau^z \rangle > 0$).

\bibliography{citations}

\begin{thebibliography}{74}%
\makeatletter
\providecommand \@ifxundefined [1]{%
 \@ifx{#1\undefined}
}%
\providecommand \@ifnum [1]{%
 \ifnum #1\expandafter \@firstoftwo
 \else \expandafter \@secondoftwo
 \fi
}%
\providecommand \@ifx [1]{%
 \ifx #1\expandafter \@firstoftwo
 \else \expandafter \@secondoftwo
 \fi
}%
\providecommand \natexlab [1]{#1}%
\providecommand \enquote  [1]{``#1''}%
\providecommand \bibnamefont  [1]{#1}%
\providecommand \bibfnamefont [1]{#1}%
\providecommand \citenamefont [1]{#1}%
\providecommand \href@noop [0]{\@secondoftwo}%
\providecommand \href [0]{\begingroup \@sanitize@url \@href}%
\providecommand \@href[1]{\@@startlink{#1}\@@href}%
\providecommand \@@href[1]{\endgroup#1\@@endlink}%
\providecommand \@sanitize@url [0]{\catcode `\\12\catcode `\$12\catcode
  `\&12\catcode `\#12\catcode `\^12\catcode `\_12\catcode `\%12\relax}%
\providecommand \@@startlink[1]{}%
\providecommand \@@endlink[0]{}%
\providecommand \url  [0]{\begingroup\@sanitize@url \@url }%
\providecommand \@url [1]{\endgroup\@href {#1}{\urlprefix }}%
\providecommand \urlprefix  [0]{URL }%
\providecommand \Eprint [0]{\href }%
\providecommand \doibase [0]{https://doi.org/}%
\providecommand \selectlanguage [0]{\@gobble}%
\providecommand \bibinfo  [0]{\@secondoftwo}%
\providecommand \bibfield  [0]{\@secondoftwo}%
\providecommand \translation [1]{[#1]}%
\providecommand \BibitemOpen [0]{}%
\providecommand \bibitemStop [0]{}%
\providecommand \bibitemNoStop [0]{.\EOS\space}%
\providecommand \EOS [0]{\spacefactor3000\relax}%
\providecommand \BibitemShut  [1]{\csname bibitem#1\endcsname}%
\let\auto@bib@innerbib\@empty
%</preamble>
\bibitem [{\citenamefont {Dagotto}(1994)}]{dagotto_94}%
  \BibitemOpen
  \bibfield  {author} {\bibinfo {author} {\bibfnamefont {E.}~\bibnamefont
  {Dagotto}},\ }\bibfield  {title} {\bibinfo {title} {Correlated electrons in
  high-temperature superconductors},\ }\href
  {https://doi.org/10.1103/RevModPhys.66.763} {\bibfield  {journal} {\bibinfo
  {journal} {Rev. Mod. Phys.}\ }\textbf {\bibinfo {volume} {66}},\ \bibinfo
  {pages} {763} (\bibinfo {year} {1994})}\BibitemShut {NoStop}%
\bibitem [{\citenamefont {Sigrist}\ and\ \citenamefont
  {Ueda}(1991)}]{sigrist_ueda_rmp_91}%
  \BibitemOpen
  \bibfield  {author} {\bibinfo {author} {\bibfnamefont {M.}~\bibnamefont
  {Sigrist}}\ and\ \bibinfo {author} {\bibfnamefont {K.}~\bibnamefont {Ueda}},\
  }\bibfield  {title} {\bibinfo {title} {Phenomenological theory of
  unconventional superconductivity},\ }\href
  {https://doi.org/10.1103/RevModPhys.63.239} {\bibfield  {journal} {\bibinfo
  {journal} {Rev. Mod. Phys.}\ }\textbf {\bibinfo {volume} {63}},\ \bibinfo
  {pages} {239} (\bibinfo {year} {1991})}\BibitemShut {NoStop}%
\bibitem [{\citenamefont {Norman}(2011)}]{norman_science_11}%
  \BibitemOpen
  \bibfield  {author} {\bibinfo {author} {\bibfnamefont {M.~R.}\ \bibnamefont
  {Norman}},\ }\bibfield  {title} {\bibinfo {title} {The challenge of
  unconventional superconductivity},\ }\href
  {https://doi.org/10.1126/science.1200181} {\bibfield  {journal} {\bibinfo
  {journal} {Science}\ }\textbf {\bibinfo {volume} {332}},\ \bibinfo {pages}
  {196} (\bibinfo {year} {2011})},\ \Eprint
  {https://arxiv.org/abs/https://www.science.org/doi/pdf/10.1126/science.1200181}
  {https://www.science.org/doi/pdf/10.1126/science.1200181} \BibitemShut
  {NoStop}%
\bibitem [{\citenamefont {Stewart}(2017)}]{stewart_aip_17}%
  \BibitemOpen
  \bibfield  {author} {\bibinfo {author} {\bibfnamefont {G.~R.}\ \bibnamefont
  {Stewart}},\ }\bibfield  {title} {\bibinfo {title} {Unconventional
  superconductivity},\ }\href {https://doi.org/10.1080/00018732.2017.1331615}
  {\bibfield  {journal} {\bibinfo  {journal} {Advances in Physics}\ }\textbf
  {\bibinfo {volume} {66}},\ \bibinfo {pages} {75} (\bibinfo {year} {2017})},\
  \Eprint {https://arxiv.org/abs/https://doi.org/10.1080/00018732.2017.1331615}
  {https://doi.org/10.1080/00018732.2017.1331615} \BibitemShut {NoStop}%
\bibitem [{\citenamefont {Bardeen}\ \emph {et~al.}(1957)\citenamefont
  {Bardeen}, \citenamefont {Cooper},\ and\ \citenamefont
  {Schrieffer}}]{bardeen_etal_pr_57}%
  \BibitemOpen
  \bibfield  {author} {\bibinfo {author} {\bibfnamefont {J.}~\bibnamefont
  {Bardeen}}, \bibinfo {author} {\bibfnamefont {L.~N.}\ \bibnamefont
  {Cooper}},\ and\ \bibinfo {author} {\bibfnamefont {J.~R.}\ \bibnamefont
  {Schrieffer}},\ }\bibfield  {title} {\bibinfo {title} {Microscopic theory of
  superconductivity},\ }\href {https://doi.org/10.1103/PhysRev.106.162}
  {\bibfield  {journal} {\bibinfo  {journal} {Phys. Rev.}\ }\textbf {\bibinfo
  {volume} {106}},\ \bibinfo {pages} {162} (\bibinfo {year}
  {1957})}\BibitemShut {NoStop}%
\bibitem [{\citenamefont {Kohn}\ and\ \citenamefont
  {Luttinger}(1965)}]{kohn_luttinger_prl_65}%
  \BibitemOpen
  \bibfield  {author} {\bibinfo {author} {\bibfnamefont {W.}~\bibnamefont
  {Kohn}}\ and\ \bibinfo {author} {\bibfnamefont {J.~M.}\ \bibnamefont
  {Luttinger}},\ }\bibfield  {title} {\bibinfo {title} {New mechanism for
  superconductivity},\ }\href {https://doi.org/10.1103/PhysRevLett.15.524}
  {\bibfield  {journal} {\bibinfo  {journal} {Phys. Rev. Lett.}\ }\textbf
  {\bibinfo {volume} {15}},\ \bibinfo {pages} {524} (\bibinfo {year}
  {1965})}\BibitemShut {NoStop}%
\bibitem [{\citenamefont {Maiti}\ and\ \citenamefont
  {Chubukov}(2013)}]{Maiti_2013}%
  \BibitemOpen
  \bibfield  {author} {\bibinfo {author} {\bibfnamefont {S.}~\bibnamefont
  {Maiti}}\ and\ \bibinfo {author} {\bibfnamefont {A.~V.}\ \bibnamefont
  {Chubukov}},\ }\bibfield  {title} {\bibinfo {title} {Superconductivity from
  repulsive interaction},\ }in\ \href {https://doi.org/10.1063/1.4818400}
  {\emph {\bibinfo {booktitle} {{AIP} Conference Proceedings}}}\ (\bibinfo
  {publisher} {{AIP}},\ \bibinfo {year} {2013})\BibitemShut {NoStop}%
\bibitem [{\citenamefont {Kagan}\ \emph {et~al.}(2014)\citenamefont {Kagan},
  \citenamefont {Val'kov}, \citenamefont {Mitskan},\ and\ \citenamefont
  {Korovushkin}}]{kagan_jetp_14}%
  \BibitemOpen
  \bibfield  {author} {\bibinfo {author} {\bibfnamefont {M.~Y.}\ \bibnamefont
  {Kagan}}, \bibinfo {author} {\bibfnamefont {V.~V.}\ \bibnamefont {Val'kov}},
  \bibinfo {author} {\bibfnamefont {V.~A.}\ \bibnamefont {Mitskan}},\ and\
  \bibinfo {author} {\bibfnamefont {M.~M.}\ \bibnamefont {Korovushkin}},\
  }\bibfield  {title} {\bibinfo {title} {The kohn-luttinger effect and
  anomalous pairing in new superconducting systems and graphene},\ }\href@noop
  {} {\bibfield  {journal} {\bibinfo  {journal} {Journal of Experimental and
  Theoretical Physics}\ }\textbf {\bibinfo {volume} {118}},\ \bibinfo {pages}
  {995} (\bibinfo {year} {2014})}\BibitemShut {NoStop}%
\bibitem [{\citenamefont {Cao}\ \emph {et~al.}(2020)\citenamefont {Cao},
  \citenamefont {Zhang}, \citenamefont {Liu}, \citenamefont {Liu},
  \citenamefont {Chen},\ and\ \citenamefont {Yang}}]{cao_etal_prl_20}%
  \BibitemOpen
  \bibfield  {author} {\bibinfo {author} {\bibfnamefont {Y.}~\bibnamefont
  {Cao}}, \bibinfo {author} {\bibfnamefont {Y.}~\bibnamefont {Zhang}}, \bibinfo
  {author} {\bibfnamefont {Y.-B.}\ \bibnamefont {Liu}}, \bibinfo {author}
  {\bibfnamefont {C.-C.}\ \bibnamefont {Liu}}, \bibinfo {author} {\bibfnamefont
  {W.-Q.}\ \bibnamefont {Chen}},\ and\ \bibinfo {author} {\bibfnamefont
  {F.}~\bibnamefont {Yang}},\ }\bibfield  {title} {\bibinfo {title}
  {Kohn-luttinger mechanism driven exotic topological superconductivity on the
  penrose lattice},\ }\href {https://doi.org/10.1103/PhysRevLett.125.017002}
  {\bibfield  {journal} {\bibinfo  {journal} {Phys. Rev. Lett.}\ }\textbf
  {\bibinfo {volume} {125}},\ \bibinfo {pages} {017002} (\bibinfo {year}
  {2020})}\BibitemShut {NoStop}%
\bibitem [{\citenamefont {Cea}\ \emph {et~al.}(2022)\citenamefont {Cea},
  \citenamefont {Pantale\'on}, \citenamefont {Phong},\ and\ \citenamefont
  {Guinea}}]{cea_etal_prb_22}%
  \BibitemOpen
  \bibfield  {author} {\bibinfo {author} {\bibfnamefont {T.}~\bibnamefont
  {Cea}}, \bibinfo {author} {\bibfnamefont {P.~A.}\ \bibnamefont
  {Pantale\'on}}, \bibinfo {author} {\bibfnamefont {V.~o.~T.}\ \bibnamefont
  {Phong}},\ and\ \bibinfo {author} {\bibfnamefont {F.}~\bibnamefont
  {Guinea}},\ }\bibfield  {title} {\bibinfo {title} {Superconductivity from
  repulsive interactions in rhombohedral trilayer graphene: A
  kohn-luttinger-like mechanism},\ }\href
  {https://doi.org/10.1103/PhysRevB.105.075432} {\bibfield  {journal} {\bibinfo
   {journal} {Phys. Rev. B}\ }\textbf {\bibinfo {volume} {105}},\ \bibinfo
  {pages} {075432} (\bibinfo {year} {2022})}\BibitemShut {NoStop}%
\bibitem [{\citenamefont {Wagner}\ \emph {et~al.}(2023)\citenamefont {Wagner},
  \citenamefont {Kwan}, \citenamefont {Bultinck}, \citenamefont {Simon},\ and\
  \citenamefont {Parameswaran}}]{wagner_etal_arxiv_23}%
  \BibitemOpen
  \bibfield  {author} {\bibinfo {author} {\bibfnamefont {G.}~\bibnamefont
  {Wagner}}, \bibinfo {author} {\bibfnamefont {Y.~H.}\ \bibnamefont {Kwan}},
  \bibinfo {author} {\bibfnamefont {N.}~\bibnamefont {Bultinck}}, \bibinfo
  {author} {\bibfnamefont {S.~H.}\ \bibnamefont {Simon}},\ and\ \bibinfo
  {author} {\bibfnamefont {S.~A.}\ \bibnamefont {Parameswaran}},\ }\href@noop
  {} {\bibinfo {title} {Superconductivity from repulsive interactions in
  bernal-stacked bilayer graphene}} (\bibinfo {year} {2023})\BibitemShut
  {NoStop}%
\bibitem [{\citenamefont {Scalapino}\ \emph {et~al.}(1986)\citenamefont
  {Scalapino}, \citenamefont {Loh},\ and\ \citenamefont
  {Hirsch}}]{scalapino_86}%
  \BibitemOpen
  \bibfield  {author} {\bibinfo {author} {\bibfnamefont {D.~J.}\ \bibnamefont
  {Scalapino}}, \bibinfo {author} {\bibfnamefont {E.}~\bibnamefont {Loh}},\
  and\ \bibinfo {author} {\bibfnamefont {J.~E.}\ \bibnamefont {Hirsch}},\
  }\bibfield  {title} {\bibinfo {title} {$d$-wave pairing near a
  spin-density-wave instability},\ }\href
  {https://doi.org/10.1103/PhysRevB.34.8190} {\bibfield  {journal} {\bibinfo
  {journal} {Phys. Rev. B}\ }\textbf {\bibinfo {volume} {34}},\ \bibinfo
  {pages} {8190} (\bibinfo {year} {1986})}\BibitemShut {NoStop}%
\bibitem [{\citenamefont {Noack}\ \emph {et~al.}(1996)\citenamefont {Noack},
  \citenamefont {Scalapino},\ and\ \citenamefont {White}}]{noack_etal_pmb_96}%
  \BibitemOpen
  \bibfield  {author} {\bibinfo {author} {\bibfnamefont {R.~M.}\ \bibnamefont
  {Noack}}, \bibinfo {author} {\bibfnamefont {D.~J.}\ \bibnamefont
  {Scalapino}},\ and\ \bibinfo {author} {\bibfnamefont {S.~R.}\ \bibnamefont
  {White}},\ }\bibfield  {title} {\bibinfo {title} {Ground-state properties of
  the two-chain hubbard ladder},\ }\href
  {https://doi.org/10.1080/01418639608240351} {\bibfield  {journal} {\bibinfo
  {journal} {Philosophical Magazine B}\ }\textbf {\bibinfo {volume} {74}},\
  \bibinfo {pages} {485} (\bibinfo {year} {1996})},\ \Eprint
  {https://arxiv.org/abs/https://doi.org/10.1080/01418639608240351}
  {https://doi.org/10.1080/01418639608240351} \BibitemShut {NoStop}%
\bibitem [{\citenamefont {Scalapino}(1999)}]{scalapino_jltp_99}%
  \BibitemOpen
  \bibfield  {author} {\bibinfo {author} {\bibfnamefont {D.~J.}\ \bibnamefont
  {Scalapino}},\ }\bibfield  {title} {\bibinfo {title} {Superconductivity and
  spin fluctuations},\ }\href {https://doi.org/10.1023/A:1022559920049}
  {\bibfield  {journal} {\bibinfo  {journal} {Journal of Low Temperature
  Physics}\ }\textbf {\bibinfo {volume} {117}},\ \bibinfo {pages} {179}
  (\bibinfo {year} {1999})}\BibitemShut {NoStop}%
\bibitem [{\citenamefont {Moriya}(2006)}]{moriya_pja_06}%
  \BibitemOpen
  \bibfield  {author} {\bibinfo {author} {\bibfnamefont {T.}~\bibnamefont
  {Moriya}},\ }\bibfield  {title} {\bibinfo {title} {Developments of the theory
  of spin fluctuations and spin fluctuation-induced superconductivity},\
  }\href@noop {} {\bibfield  {journal} {\bibinfo  {journal} {Proc Jpn Acad Ser
  B Phys Biol Sci}\ }\textbf {\bibinfo {volume} {82}},\ \bibinfo {pages} {1}
  (\bibinfo {year} {2006})}\BibitemShut {NoStop}%
\bibitem [{\citenamefont {Chang}\ \emph {et~al.}(2020)\citenamefont {Chang},
  \citenamefont {Zhao},\ and\ \citenamefont {Ding}}]{chang_etal_epjb_20}%
  \BibitemOpen
  \bibfield  {author} {\bibinfo {author} {\bibfnamefont {J.}~\bibnamefont
  {Chang}}, \bibinfo {author} {\bibfnamefont {J.}~\bibnamefont {Zhao}},\ and\
  \bibinfo {author} {\bibfnamefont {Y.}~\bibnamefont {Ding}},\ }\bibfield
  {title} {\bibinfo {title} {Hund-heisenberg model in superconducting
  infinite-layer nickelates},\ }\href
  {https://doi.org/10.1140/epjb/e2020-10343-7} {\bibfield  {journal} {\bibinfo
  {journal} {The European Physical Journal B}\ }\textbf {\bibinfo {volume}
  {93}},\ \bibinfo {pages} {220} (\bibinfo {year} {2020})}\BibitemShut
  {NoStop}%
\bibitem [{\citenamefont {Vafek}\ and\ \citenamefont
  {Chubukov}(2017)}]{vafek_chubokov_prl_17}%
  \BibitemOpen
  \bibfield  {author} {\bibinfo {author} {\bibfnamefont {O.}~\bibnamefont
  {Vafek}}\ and\ \bibinfo {author} {\bibfnamefont {A.~V.}\ \bibnamefont
  {Chubukov}},\ }\bibfield  {title} {\bibinfo {title} {Hund interaction,
  spin-orbit coupling, and the mechanism of superconductivity in strongly
  hole-doped iron pnictides},\ }\href
  {https://doi.org/10.1103/PhysRevLett.118.087003} {\bibfield  {journal}
  {\bibinfo  {journal} {Phys. Rev. Lett.}\ }\textbf {\bibinfo {volume} {118}},\
  \bibinfo {pages} {087003} (\bibinfo {year} {2017})}\BibitemShut {NoStop}%
\bibitem [{\citenamefont {Cheung}\ and\ \citenamefont
  {Agterberg}(2019)}]{cheung_agterberg_prb_19}%
  \BibitemOpen
  \bibfield  {author} {\bibinfo {author} {\bibfnamefont {A.~K.~C.}\
  \bibnamefont {Cheung}}\ and\ \bibinfo {author} {\bibfnamefont {D.~F.}\
  \bibnamefont {Agterberg}},\ }\bibfield  {title} {\bibinfo {title}
  {Superconductivity in the presence of spin-orbit interactions stabilized by
  hund coupling},\ }\href {https://doi.org/10.1103/PhysRevB.99.024516}
  {\bibfield  {journal} {\bibinfo  {journal} {Phys. Rev. B}\ }\textbf {\bibinfo
  {volume} {99}},\ \bibinfo {pages} {024516} (\bibinfo {year}
  {2019})}\BibitemShut {NoStop}%
\bibitem [{\citenamefont {Karakonstantakis}\ \emph {et~al.}(2011)\citenamefont
  {Karakonstantakis}, \citenamefont {Berg}, \citenamefont {White},\ and\
  \citenamefont {Kivelson}}]{karakonstankis_etal_prb_11}%
  \BibitemOpen
  \bibfield  {author} {\bibinfo {author} {\bibfnamefont {G.}~\bibnamefont
  {Karakonstantakis}}, \bibinfo {author} {\bibfnamefont {E.}~\bibnamefont
  {Berg}}, \bibinfo {author} {\bibfnamefont {S.~R.}\ \bibnamefont {White}},\
  and\ \bibinfo {author} {\bibfnamefont {S.~A.}\ \bibnamefont {Kivelson}},\
  }\bibfield  {title} {\bibinfo {title} {Enhanced pairing in the checkerboard
  hubbard ladder},\ }\href {https://doi.org/10.1103/PhysRevB.83.054508}
  {\bibfield  {journal} {\bibinfo  {journal} {Phys. Rev. B}\ }\textbf {\bibinfo
  {volume} {83}},\ \bibinfo {pages} {054508} (\bibinfo {year}
  {2011})}\BibitemShut {NoStop}%
\bibitem [{\citenamefont {Ehlers}\ \emph {et~al.}(2017)\citenamefont {Ehlers},
  \citenamefont {White},\ and\ \citenamefont {Noack}}]{ehlers_eal_prb_17}%
  \BibitemOpen
  \bibfield  {author} {\bibinfo {author} {\bibfnamefont {G.}~\bibnamefont
  {Ehlers}}, \bibinfo {author} {\bibfnamefont {S.~R.}\ \bibnamefont {White}},\
  and\ \bibinfo {author} {\bibfnamefont {R.~M.}\ \bibnamefont {Noack}},\
  }\bibfield  {title} {\bibinfo {title} {Hybrid-space density matrix
  renormalization group study of the doped two-dimensional hubbard model},\
  }\href {https://doi.org/10.1103/PhysRevB.95.125125} {\bibfield  {journal}
  {\bibinfo  {journal} {Phys. Rev. B}\ }\textbf {\bibinfo {volume} {95}},\
  \bibinfo {pages} {125125} (\bibinfo {year} {2017})}\BibitemShut {NoStop}%
\bibitem [{\citenamefont {Jiang}\ and\ \citenamefont
  {Devereaux}(2019)}]{jiang_devereaux_science_19}%
  \BibitemOpen
  \bibfield  {author} {\bibinfo {author} {\bibfnamefont {H.-C.}\ \bibnamefont
  {Jiang}}\ and\ \bibinfo {author} {\bibfnamefont {T.~P.}\ \bibnamefont
  {Devereaux}},\ }\bibfield  {title} {\bibinfo {title} {Superconductivity in
  the doped hubbard model and its interplay with next-nearest hopping
  <i>t</i>\&\#x2032;},\ }\href {https://doi.org/10.1126/science.aal5304}
  {\bibfield  {journal} {\bibinfo  {journal} {Science}\ }\textbf {\bibinfo
  {volume} {365}},\ \bibinfo {pages} {1424} (\bibinfo {year} {2019})},\ \Eprint
  {https://arxiv.org/abs/https://www.science.org/doi/pdf/10.1126/science.aal5304}
  {https://www.science.org/doi/pdf/10.1126/science.aal5304} \BibitemShut
  {NoStop}%
\bibitem [{\citenamefont {Jiang}\ and\ \citenamefont
  {Devereaux}(2023)}]{jiang_devereaux_fron_23}%
  \BibitemOpen
  \bibfield  {author} {\bibinfo {author} {\bibfnamefont {H.-C.}\ \bibnamefont
  {Jiang}}\ and\ \bibinfo {author} {\bibfnamefont {T.~P.}\ \bibnamefont
  {Devereaux}},\ }\bibfield  {title} {\bibinfo {title} {Pair density wave and
  superconductivity in a kinetically frustrated doped emery model on a square
  lattice},\ }\bibfield  {journal} {\bibinfo  {journal} {Frontiers in
  Electronic Materials}\ }\textbf {\bibinfo {volume} {3}},\ \href
  {https://doi.org/10.3389/femat.2023.1323404} {10.3389/femat.2023.1323404}
  (\bibinfo {year} {2023})\BibitemShut {NoStop}%
\bibitem [{\citenamefont {Jiang}(2023)}]{jiang_prb_23}%
  \BibitemOpen
  \bibfield  {author} {\bibinfo {author} {\bibfnamefont {H.-C.}\ \bibnamefont
  {Jiang}},\ }\bibfield  {title} {\bibinfo {title} {Pair density wave in the
  doped three-band hubbard model on two-leg square cylinders},\ }\href
  {https://doi.org/10.1103/PhysRevB.107.214504} {\bibfield  {journal} {\bibinfo
   {journal} {Phys. Rev. B}\ }\textbf {\bibinfo {volume} {107}},\ \bibinfo
  {pages} {214504} (\bibinfo {year} {2023})}\BibitemShut {NoStop}%
\bibitem [{\citenamefont {Georges}\ \emph {et~al.}(2013)\citenamefont
  {Georges}, \citenamefont {Medici},\ and\ \citenamefont
  {Mravlje}}]{georges_etal_arcmp_13}%
  \BibitemOpen
  \bibfield  {author} {\bibinfo {author} {\bibfnamefont {A.}~\bibnamefont
  {Georges}}, \bibinfo {author} {\bibfnamefont {L.~d.}\ \bibnamefont
  {Medici}},\ and\ \bibinfo {author} {\bibfnamefont {J.}~\bibnamefont
  {Mravlje}},\ }\bibfield  {title} {\bibinfo {title} {Strong correlations from
  hund’s coupling},\ }\href
  {https://doi.org/10.1146/annurev-conmatphys-020911-125045} {\bibfield
  {journal} {\bibinfo  {journal} {Annual Review of Condensed Matter Physics}\
  }\textbf {\bibinfo {volume} {4}},\ \bibinfo {pages} {137} (\bibinfo {year}
  {2013})},\ \Eprint
  {https://arxiv.org/abs/https://doi.org/10.1146/annurev-conmatphys-020911-125045}
  {https://doi.org/10.1146/annurev-conmatphys-020911-125045} \BibitemShut
  {NoStop}%
\bibitem [{\citenamefont {Nomura}\ \emph {et~al.}(2015)\citenamefont {Nomura},
  \citenamefont {Sakai}, \citenamefont {Capone},\ and\ \citenamefont
  {Arita}}]{nomura_etal_sa_15}%
  \BibitemOpen
  \bibfield  {author} {\bibinfo {author} {\bibfnamefont {Y.}~\bibnamefont
  {Nomura}}, \bibinfo {author} {\bibfnamefont {S.}~\bibnamefont {Sakai}},
  \bibinfo {author} {\bibfnamefont {M.}~\bibnamefont {Capone}},\ and\ \bibinfo
  {author} {\bibfnamefont {R.}~\bibnamefont {Arita}},\ }\bibfield  {title}
  {\bibinfo {title} {Unified understanding of superconductivity and mott
  transition in alkali-doped fullerides from first principles},\ }\href
  {https://doi.org/10.1126/sciadv.1500568} {\bibfield  {journal} {\bibinfo
  {journal} {Science Advances}\ }\textbf {\bibinfo {volume} {1}},\ \bibinfo
  {pages} {e1500568} (\bibinfo {year} {2015})},\ \Eprint
  {https://arxiv.org/abs/https://www.science.org/doi/pdf/10.1126/sciadv.1500568}
  {https://www.science.org/doi/pdf/10.1126/sciadv.1500568} \BibitemShut
  {NoStop}%
\bibitem [{\citenamefont {Pavarini}\ \emph {et~al.}(2017)\citenamefont
  {Pavarini}, \citenamefont {Koch}, \citenamefont {Scalettar},\ and\
  \citenamefont {Martin}}]{pavarini_book_17}%
  \BibitemOpen
  \bibinfo {editor} {\bibfnamefont {E.}~\bibnamefont {Pavarini}}, \bibinfo
  {editor} {\bibfnamefont {E.}~\bibnamefont {Koch}}, \bibinfo {editor}
  {\bibfnamefont {R.}~\bibnamefont {Scalettar}},\ and\ \bibinfo {editor}
  {\bibfnamefont {R.}~\bibnamefont {Martin}},\ eds.,\ \href
  {https://juser.fz-juelich.de/record/837488} {\emph {\bibinfo {title} {{T}he
  {P}hysics of {C}orrelated {I}nsulators, {M}etals, and {S}uperconductors}}},\
  \bibinfo {series} {Schriften des Forschungszentrums Jülich. Reihe Modeling
  and Simulation}, Vol.~\bibinfo {volume} {7}\ (\bibinfo  {publisher}
  {Forschungszentrum Jülich GmbH Zentralbibliothek, Verlag},\ \bibinfo
  {address} {Jülich},\ \bibinfo {year} {2017})\ p.\ \bibinfo {pages} {450
  S.}\BibitemShut {Stop}%
\bibitem [{\citenamefont {Fanfarillo}\ \emph {et~al.}(2020)\citenamefont
  {Fanfarillo}, \citenamefont {Valli},\ and\ \citenamefont
  {Capone}}]{fanfarillo_etal_prl_20}%
  \BibitemOpen
  \bibfield  {author} {\bibinfo {author} {\bibfnamefont {L.}~\bibnamefont
  {Fanfarillo}}, \bibinfo {author} {\bibfnamefont {A.}~\bibnamefont {Valli}},\
  and\ \bibinfo {author} {\bibfnamefont {M.}~\bibnamefont {Capone}},\
  }\bibfield  {title} {\bibinfo {title} {Synergy between hund-driven
  correlations and boson-mediated superconductivity},\ }\href
  {https://doi.org/10.1103/PhysRevLett.125.177001} {\bibfield  {journal}
  {\bibinfo  {journal} {Phys. Rev. Lett.}\ }\textbf {\bibinfo {volume} {125}},\
  \bibinfo {pages} {177001} (\bibinfo {year} {2020})}\BibitemShut {NoStop}%
\bibitem [{\citenamefont {Liu}\ \emph {et~al.}(2024)\citenamefont {Liu},
  \citenamefont {Peng}, \citenamefont {Huang}, \citenamefont {Moritz},
  \citenamefont {Jia},\ and\ \citenamefont {Devereaux}}]{liu_npj_qmat_24}%
  \BibitemOpen
  \bibfield  {author} {\bibinfo {author} {\bibfnamefont {F.}~\bibnamefont
  {Liu}}, \bibinfo {author} {\bibfnamefont {C.}~\bibnamefont {Peng}}, \bibinfo
  {author} {\bibfnamefont {E.~W.}\ \bibnamefont {Huang}}, \bibinfo {author}
  {\bibfnamefont {B.}~\bibnamefont {Moritz}}, \bibinfo {author} {\bibfnamefont
  {C.}~\bibnamefont {Jia}},\ and\ \bibinfo {author} {\bibfnamefont {T.~P.}\
  \bibnamefont {Devereaux}},\ }\bibfield  {title} {\bibinfo {title} {Emergence
  of antiferromagnetic correlations and kondolike features in a model for
  infinite layer nickelates},\ }\href@noop {} {\bibfield  {journal} {\bibinfo
  {journal} {npj Quantum Materials}\ }\textbf {\bibinfo {volume} {9}},\
  \bibinfo {pages} {49} (\bibinfo {year} {2024})}\BibitemShut {NoStop}%
\bibitem [{\citenamefont {Tsai}\ and\ \citenamefont
  {Kivelson}(2006)}]{tsai_kivelson_prb_06}%
  \BibitemOpen
  \bibfield  {author} {\bibinfo {author} {\bibfnamefont {W.-F.}\ \bibnamefont
  {Tsai}}\ and\ \bibinfo {author} {\bibfnamefont {S.~A.}\ \bibnamefont
  {Kivelson}},\ }\bibfield  {title} {\bibinfo {title} {Superconductivity in
  inhomogeneous hubbard models},\ }\href
  {https://doi.org/10.1103/PhysRevB.73.214510} {\bibfield  {journal} {\bibinfo
  {journal} {Phys. Rev. B}\ }\textbf {\bibinfo {volume} {73}},\ \bibinfo
  {pages} {214510} (\bibinfo {year} {2006})}\BibitemShut {NoStop}%
\bibitem [{\citenamefont {Slagle}\ and\ \citenamefont
  {Kim}(2019)}]{slagle_kim_scipost_19}%
  \BibitemOpen
  \bibfield  {author} {\bibinfo {author} {\bibfnamefont {K.}~\bibnamefont
  {Slagle}}\ and\ \bibinfo {author} {\bibfnamefont {Y.~B.}\ \bibnamefont
  {Kim}},\ }\bibfield  {title} {\bibinfo {title} {{A simple mechanism for
  unconventional superconductivity in a repulsive fermion model}},\ }\href
  {https://doi.org/10.21468/SciPostPhys.6.2.016} {\bibfield  {journal}
  {\bibinfo  {journal} {SciPost Phys.}\ }\textbf {\bibinfo {volume} {6}},\
  \bibinfo {pages} {016} (\bibinfo {year} {2019})}\BibitemShut {NoStop}%
\bibitem [{\citenamefont {Crépel}\ and\ \citenamefont
  {Fu}(2021)}]{crepel_fu_scad_21}%
  \BibitemOpen
  \bibfield  {author} {\bibinfo {author} {\bibfnamefont {V.}~\bibnamefont
  {Crépel}}\ and\ \bibinfo {author} {\bibfnamefont {L.}~\bibnamefont {Fu}},\
  }\bibfield  {title} {\bibinfo {title} {New mechanism and exact theory of
  superconductivity from strong repulsive interaction},\ }\href
  {https://doi.org/10.1126/sciadv.abh2233} {\bibfield  {journal} {\bibinfo
  {journal} {Science Advances}\ }\textbf {\bibinfo {volume} {7}},\ \bibinfo
  {pages} {eabh2233} (\bibinfo {year} {2021})},\ \Eprint
  {https://arxiv.org/abs/https://www.science.org/doi/pdf/10.1126/sciadv.abh2233}
  {https://www.science.org/doi/pdf/10.1126/sciadv.abh2233} \BibitemShut
  {NoStop}%
\bibitem [{\citenamefont {Cr\'epel}\ \emph {et~al.}(2022)\citenamefont
  {Cr\'epel}, \citenamefont {Cea}, \citenamefont {Fu},\ and\ \citenamefont
  {Guinea}}]{crepel_etal_prb_22}%
  \BibitemOpen
  \bibfield  {author} {\bibinfo {author} {\bibfnamefont {V.}~\bibnamefont
  {Cr\'epel}}, \bibinfo {author} {\bibfnamefont {T.}~\bibnamefont {Cea}},
  \bibinfo {author} {\bibfnamefont {L.}~\bibnamefont {Fu}},\ and\ \bibinfo
  {author} {\bibfnamefont {F.}~\bibnamefont {Guinea}},\ }\bibfield  {title}
  {\bibinfo {title} {Unconventional superconductivity due to interband
  polarization},\ }\href {https://doi.org/10.1103/PhysRevB.105.094506}
  {\bibfield  {journal} {\bibinfo  {journal} {Phys. Rev. B}\ }\textbf {\bibinfo
  {volume} {105}},\ \bibinfo {pages} {094506} (\bibinfo {year}
  {2022})}\BibitemShut {NoStop}%
\bibitem [{\citenamefont {Crépel}\ and\ \citenamefont
  {Fu}(2022)}]{crepel_fu_proc_22}%
  \BibitemOpen
  \bibfield  {author} {\bibinfo {author} {\bibfnamefont {V.}~\bibnamefont
  {Crépel}}\ and\ \bibinfo {author} {\bibfnamefont {L.}~\bibnamefont {Fu}},\
  }\bibfield  {title} {\bibinfo {title} {Spin-triplet superconductivity from
  excitonic effect in doped insulators},\ }\href
  {https://doi.org/10.1073/pnas.2117735119} {\bibfield  {journal} {\bibinfo
  {journal} {Proceedings of the National Academy of Sciences}\ }\textbf
  {\bibinfo {volume} {119}},\ \bibinfo {pages} {e2117735119} (\bibinfo {year}
  {2022})},\ \Eprint
  {https://arxiv.org/abs/https://www.pnas.org/doi/pdf/10.1073/pnas.2117735119}
  {https://www.pnas.org/doi/pdf/10.1073/pnas.2117735119} \BibitemShut {NoStop}%
\bibitem [{\citenamefont {He}\ \emph {et~al.}(2023)\citenamefont {He},
  \citenamefont {Yang}, \citenamefont {Hauck}, \citenamefont {Bergholtz},\ and\
  \citenamefont {Kennes}}]{he_etal_prr_23}%
  \BibitemOpen
  \bibfield  {author} {\bibinfo {author} {\bibfnamefont {Y.}~\bibnamefont
  {He}}, \bibinfo {author} {\bibfnamefont {K.}~\bibnamefont {Yang}}, \bibinfo
  {author} {\bibfnamefont {J.~B.}\ \bibnamefont {Hauck}}, \bibinfo {author}
  {\bibfnamefont {E.~J.}\ \bibnamefont {Bergholtz}},\ and\ \bibinfo {author}
  {\bibfnamefont {D.~M.}\ \bibnamefont {Kennes}},\ }\bibfield  {title}
  {\bibinfo {title} {Superconductivity of repulsive spinless fermions with
  sublattice potentials},\ }\href
  {https://doi.org/10.1103/PhysRevResearch.5.L012009} {\bibfield  {journal}
  {\bibinfo  {journal} {Phys. Rev. Res.}\ }\textbf {\bibinfo {volume} {5}},\
  \bibinfo {pages} {L012009} (\bibinfo {year} {2023})}\BibitemShut {NoStop}%
\bibitem [{\citenamefont {Zhang}\ \emph {et~al.}(2020)\citenamefont {Zhang},
  \citenamefont {Yang},\ and\ \citenamefont {Zhang}}]{zhang_etal_prb_20}%
  \BibitemOpen
  \bibfield  {author} {\bibinfo {author} {\bibfnamefont {G.-M.}\ \bibnamefont
  {Zhang}}, \bibinfo {author} {\bibfnamefont {Y.-f.}\ \bibnamefont {Yang}},\
  and\ \bibinfo {author} {\bibfnamefont {F.-C.}\ \bibnamefont {Zhang}},\
  }\bibfield  {title} {\bibinfo {title} {Self-doped mott insulator for parent
  compounds of nickelate superconductors},\ }\href
  {https://doi.org/10.1103/PhysRevB.101.020501} {\bibfield  {journal} {\bibinfo
   {journal} {Phys. Rev. B}\ }\textbf {\bibinfo {volume} {101}},\ \bibinfo
  {pages} {020501} (\bibinfo {year} {2020})}\BibitemShut {NoStop}%
\bibitem [{\citenamefont {Yang}\ and\ \citenamefont
  {Zhang}(2022)}]{yang_zhang_fron_22}%
  \BibitemOpen
  \bibfield  {author} {\bibinfo {author} {\bibfnamefont {Y.-f.}\ \bibnamefont
  {Yang}}\ and\ \bibinfo {author} {\bibfnamefont {G.-M.}\ \bibnamefont
  {Zhang}},\ }\bibfield  {title} {\bibinfo {title} {Self-doping and the
  mott-kondo scenario for infinite-layer nickelate superconductors},\
  }\bibfield  {journal} {\bibinfo  {journal} {Frontiers in Physics}\ }\textbf
  {\bibinfo {volume} {9}},\ \href {https://doi.org/10.3389/fphy.2021.801236}
  {10.3389/fphy.2021.801236} (\bibinfo {year} {2022})\BibitemShut {NoStop}%
\bibitem [{\citenamefont {Chen}\ \emph
  {et~al.}(2022{\natexlab{a}})\citenamefont {Chen}, \citenamefont {Jiang},
  \citenamefont {Si}, \citenamefont {Lu},\ and\ \citenamefont
  {Zhong}}]{chen_etal_prb_22}%
  \BibitemOpen
  \bibfield  {author} {\bibinfo {author} {\bibfnamefont {D.}~\bibnamefont
  {Chen}}, \bibinfo {author} {\bibfnamefont {P.}~\bibnamefont {Jiang}},
  \bibinfo {author} {\bibfnamefont {L.}~\bibnamefont {Si}}, \bibinfo {author}
  {\bibfnamefont {Y.}~\bibnamefont {Lu}},\ and\ \bibinfo {author}
  {\bibfnamefont {Z.}~\bibnamefont {Zhong}},\ }\bibfield  {title} {\bibinfo
  {title} {Magnetism in doped infinite-layer ${\mathrm{ndnio}}_{2}$ studied by
  combined density functional theory and dynamical mean-field theory},\ }\href
  {https://doi.org/10.1103/PhysRevB.106.045105} {\bibfield  {journal} {\bibinfo
   {journal} {Phys. Rev. B}\ }\textbf {\bibinfo {volume} {106}},\ \bibinfo
  {pages} {045105} (\bibinfo {year} {2022}{\natexlab{a}})}\BibitemShut
  {NoStop}%
\bibitem [{\citenamefont {Chen}\ \emph
  {et~al.}(2022{\natexlab{b}})\citenamefont {Chen}, \citenamefont {Hampel},
  \citenamefont {Karp}, \citenamefont {Lechermann},\ and\ \citenamefont
  {Millis}}]{chen_etal_fron_22}%
  \BibitemOpen
  \bibfield  {author} {\bibinfo {author} {\bibfnamefont {H.}~\bibnamefont
  {Chen}}, \bibinfo {author} {\bibfnamefont {A.}~\bibnamefont {Hampel}},
  \bibinfo {author} {\bibfnamefont {J.}~\bibnamefont {Karp}}, \bibinfo {author}
  {\bibfnamefont {F.}~\bibnamefont {Lechermann}},\ and\ \bibinfo {author}
  {\bibfnamefont {A.~J.}\ \bibnamefont {Millis}},\ }\bibfield  {title}
  {\bibinfo {title} {Dynamical mean field studies of infinite layer nickelates:
  Physics results and methodological implications},\ }\bibfield  {journal}
  {\bibinfo  {journal} {Frontiers in Physics}\ }\textbf {\bibinfo {volume}
  {10}},\ \href {https://doi.org/10.3389/fphy.2022.835942}
  {10.3389/fphy.2022.835942} (\bibinfo {year} {2022}{\natexlab{b}})\BibitemShut
  {NoStop}%
\bibitem [{\citenamefont {Fowlie}\ \emph {et~al.}(2022)\citenamefont {Fowlie},
  \citenamefont {Hadjimichael}, \citenamefont {Martins}, \citenamefont {Li},
  \citenamefont {Osada}, \citenamefont {Wang}, \citenamefont {Lee},
  \citenamefont {Lee}, \citenamefont {Salman}, \citenamefont {Prokscha},
  \citenamefont {Triscone}, \citenamefont {Hwang},\ and\ \citenamefont
  {Suter}}]{fowlie_etal_natphys_22}%
  \BibitemOpen
  \bibfield  {author} {\bibinfo {author} {\bibfnamefont {J.}~\bibnamefont
  {Fowlie}}, \bibinfo {author} {\bibfnamefont {M.}~\bibnamefont
  {Hadjimichael}}, \bibinfo {author} {\bibfnamefont {M.~M.}\ \bibnamefont
  {Martins}}, \bibinfo {author} {\bibfnamefont {D.}~\bibnamefont {Li}},
  \bibinfo {author} {\bibfnamefont {M.}~\bibnamefont {Osada}}, \bibinfo
  {author} {\bibfnamefont {B.~Y.}\ \bibnamefont {Wang}}, \bibinfo {author}
  {\bibfnamefont {K.}~\bibnamefont {Lee}}, \bibinfo {author} {\bibfnamefont
  {Y.}~\bibnamefont {Lee}}, \bibinfo {author} {\bibfnamefont {Z.}~\bibnamefont
  {Salman}}, \bibinfo {author} {\bibfnamefont {T.}~\bibnamefont {Prokscha}},
  \bibinfo {author} {\bibfnamefont {J.-M.}\ \bibnamefont {Triscone}}, \bibinfo
  {author} {\bibfnamefont {H.~Y.}\ \bibnamefont {Hwang}},\ and\ \bibinfo
  {author} {\bibfnamefont {A.}~\bibnamefont {Suter}},\ }\bibfield  {title}
  {\bibinfo {title} {Intrinsic magnetism in superconducting infinite-layer
  nickelates},\ }\href {https://doi.org/10.1038/s41567-022-01684-y} {\bibfield
  {journal} {\bibinfo  {journal} {Nature Physics}\ }\textbf {\bibinfo {volume}
  {18}},\ \bibinfo {pages} {1043} (\bibinfo {year} {2022})}\BibitemShut
  {NoStop}%
\bibitem [{\citenamefont {Nomura}\ and\ \citenamefont
  {Arita}(2022)}]{nomura_arita_iop_22}%
  \BibitemOpen
  \bibfield  {author} {\bibinfo {author} {\bibfnamefont {Y.}~\bibnamefont
  {Nomura}}\ and\ \bibinfo {author} {\bibfnamefont {R.}~\bibnamefont {Arita}},\
  }\bibfield  {title} {\bibinfo {title} {Superconductivity in infinite-layer
  nickelates},\ }\href {https://doi.org/10.1088/1361-6633/ac5a60} {\bibfield
  {journal} {\bibinfo  {journal} {Reports on Progress in Physics}\ }\textbf
  {\bibinfo {volume} {85}},\ \bibinfo {pages} {052501} (\bibinfo {year}
  {2022})}\BibitemShut {NoStop}%
\bibitem [{\citenamefont {Park}\ \emph {et~al.}(2021)\citenamefont {Park},
  \citenamefont {Cao}, \citenamefont {Watanabe}, \citenamefont {Taniguchi},\
  and\ \citenamefont {Jarillo-Herrero}}]{park_etal_nature_21}%
  \BibitemOpen
  \bibfield  {author} {\bibinfo {author} {\bibfnamefont {J.~M.}\ \bibnamefont
  {Park}}, \bibinfo {author} {\bibfnamefont {Y.}~\bibnamefont {Cao}}, \bibinfo
  {author} {\bibfnamefont {K.}~\bibnamefont {Watanabe}}, \bibinfo {author}
  {\bibfnamefont {T.}~\bibnamefont {Taniguchi}},\ and\ \bibinfo {author}
  {\bibfnamefont {P.}~\bibnamefont {Jarillo-Herrero}},\ }\bibfield  {title}
  {\bibinfo {title} {Tunable strongly coupled superconductivity in magic-angle
  twisted trilayer graphene},\ }\href
  {https://doi.org/10.1038/s41586-021-03192-0} {\bibfield  {journal} {\bibinfo
  {journal} {Nature}\ }\textbf {\bibinfo {volume} {590}},\ \bibinfo {pages}
  {249} (\bibinfo {year} {2021})}\BibitemShut {NoStop}%
\bibitem [{\citenamefont {Bodensiek}\ \emph {et~al.}(2010)\citenamefont
  {Bodensiek}, \citenamefont {Pruschke},\ and\ \citenamefont
  {Žitko}}]{bodensiek_etal_jop_10}%
  \BibitemOpen
  \bibfield  {author} {\bibinfo {author} {\bibfnamefont {O.}~\bibnamefont
  {Bodensiek}}, \bibinfo {author} {\bibfnamefont {T.}~\bibnamefont
  {Pruschke}},\ and\ \bibinfo {author} {\bibfnamefont {R.}~\bibnamefont
  {Žitko}},\ }\bibfield  {title} {\bibinfo {title} {Superconductivity in the
  kondo lattice model},\ }\href
  {https://doi.org/10.1088/1742-6596/200/1/012162} {\bibfield  {journal}
  {\bibinfo  {journal} {Journal of Physics: Conference Series}\ }\textbf
  {\bibinfo {volume} {200}},\ \bibinfo {pages} {012162} (\bibinfo {year}
  {2010})}\BibitemShut {NoStop}%
\bibitem [{\citenamefont {Lynn}\ \emph {et~al.}(2023)\citenamefont {Lynn},
  \citenamefont {Madhavan},\ and\ \citenamefont {Jiao}}]{lynn_etal_fron_23}%
  \BibitemOpen
  \bibfield  {author} {\bibinfo {author} {\bibfnamefont {J.~W.}\ \bibnamefont
  {Lynn}}, \bibinfo {author} {\bibfnamefont {V.}~\bibnamefont {Madhavan}},\
  and\ \bibinfo {author} {\bibfnamefont {L.}~\bibnamefont {Jiao}},\ }\bibfield
  {title} {\bibinfo {title} {Editorial: New heavy fermion superconductors},\
  }\bibfield  {journal} {\bibinfo  {journal} {Frontiers in Electronic
  Materials}\ }\textbf {\bibinfo {volume} {2}},\ \href
  {https://doi.org/10.3389/femat.2022.1120381} {10.3389/femat.2022.1120381}
  (\bibinfo {year} {2023})\BibitemShut {NoStop}%
\bibitem [{\citenamefont {Schrieffer}\ and\ \citenamefont
  {Wolff}(1966)}]{schrieffer_66}%
  \BibitemOpen
  \bibfield  {author} {\bibinfo {author} {\bibfnamefont {J.~R.}\ \bibnamefont
  {Schrieffer}}\ and\ \bibinfo {author} {\bibfnamefont {P.~A.}\ \bibnamefont
  {Wolff}},\ }\bibfield  {title} {\bibinfo {title} {Relation between the
  anderson and kondo hamiltonians},\ }\href
  {https://doi.org/10.1103/PhysRev.149.491} {\bibfield  {journal} {\bibinfo
  {journal} {Phys. Rev.}\ }\textbf {\bibinfo {volume} {149}},\ \bibinfo {pages}
  {491} (\bibinfo {year} {1966})}\BibitemShut {NoStop}%
\bibitem [{\citenamefont {Coleman}(2015)}]{coleman}%
  \BibitemOpen
  \bibfield  {author} {\bibinfo {author} {\bibfnamefont {P.}~\bibnamefont
  {Coleman}},\ }\href@noop {} {\bibinfo {title} {Heavy fermions and the kondo
  lattice: a 21st century perspective}} (\bibinfo {year} {2015}),\ \Eprint
  {https://arxiv.org/abs/1509.05769} {arXiv:1509.05769 [cond-mat.str-el]}
  \BibitemShut {NoStop}%
\bibitem [{\citenamefont {Zhao}\ \emph {et~al.}(2023)\citenamefont {Zhao},
  \citenamefont {Shen}, \citenamefont {Tao}, \citenamefont {Han}, \citenamefont
  {Kang}, \citenamefont {Watanabe}, \citenamefont {Taniguchi}, \citenamefont
  {Mak},\ and\ \citenamefont {Shan}}]{zhao_etal_nature_23}%
  \BibitemOpen
  \bibfield  {author} {\bibinfo {author} {\bibfnamefont {W.}~\bibnamefont
  {Zhao}}, \bibinfo {author} {\bibfnamefont {B.}~\bibnamefont {Shen}}, \bibinfo
  {author} {\bibfnamefont {Z.}~\bibnamefont {Tao}}, \bibinfo {author}
  {\bibfnamefont {Z.}~\bibnamefont {Han}}, \bibinfo {author} {\bibfnamefont
  {K.}~\bibnamefont {Kang}}, \bibinfo {author} {\bibfnamefont {K.}~\bibnamefont
  {Watanabe}}, \bibinfo {author} {\bibfnamefont {T.}~\bibnamefont {Taniguchi}},
  \bibinfo {author} {\bibfnamefont {K.~F.}\ \bibnamefont {Mak}},\ and\ \bibinfo
  {author} {\bibfnamefont {J.}~\bibnamefont {Shan}},\ }\bibfield  {title}
  {\bibinfo {title} {Gate-tunable heavy fermions in a moir{\'{e}} kondo
  lattice},\ }\href {https://doi.org/10.1038/s41586-023-05800-7} {\bibfield
  {journal} {\bibinfo  {journal} {Nature}\ }\textbf {\bibinfo {volume} {616}},\
  \bibinfo {pages} {61} (\bibinfo {year} {2023})}\BibitemShut {NoStop}%
\bibitem [{\citenamefont {Giamarchi}(2003)}]{giamarchi_book_03}%
  \BibitemOpen
  \bibfield  {author} {\bibinfo {author} {\bibfnamefont {T.}~\bibnamefont
  {Giamarchi}},\ }\href@noop {} {\emph {\bibinfo {title} {Quantum Physics in
  One Dimension}}},\ International Series of Monographs on Physics\ (\bibinfo
  {publisher} {Oxford University Press},\ \bibinfo {address} {Oxford},\
  \bibinfo {year} {2003})\BibitemShut {NoStop}%
\bibitem [{\citenamefont {Patel}\ \emph {et~al.}(2017)\citenamefont {Patel},
  \citenamefont {Nocera}, \citenamefont {Alvarez}, \citenamefont {Moreo},\ and\
  \citenamefont {Dagotto}}]{patel_etal_prb_17}%
  \BibitemOpen
  \bibfield  {author} {\bibinfo {author} {\bibfnamefont {N.~D.}\ \bibnamefont
  {Patel}}, \bibinfo {author} {\bibfnamefont {A.}~\bibnamefont {Nocera}},
  \bibinfo {author} {\bibfnamefont {G.}~\bibnamefont {Alvarez}}, \bibinfo
  {author} {\bibfnamefont {A.}~\bibnamefont {Moreo}},\ and\ \bibinfo {author}
  {\bibfnamefont {E.}~\bibnamefont {Dagotto}},\ }\bibfield  {title} {\bibinfo
  {title} {Pairing tendencies in a two-orbital hubbard model in one
  dimension},\ }\href {https://doi.org/10.1103/PhysRevB.96.024520} {\bibfield
  {journal} {\bibinfo  {journal} {Phys. Rev. B}\ }\textbf {\bibinfo {volume}
  {96}},\ \bibinfo {pages} {024520} (\bibinfo {year} {2017})}\BibitemShut
  {NoStop}%
\bibitem [{\citenamefont {Nocera}\ \emph {et~al.}(2018)\citenamefont {Nocera},
  \citenamefont {Wang}, \citenamefont {Patel}, \citenamefont {Alvarez},
  \citenamefont {Maier}, \citenamefont {Dagotto},\ and\ \citenamefont
  {Johnston}}]{nocera_etal_prb_18}%
  \BibitemOpen
  \bibfield  {author} {\bibinfo {author} {\bibfnamefont {A.}~\bibnamefont
  {Nocera}}, \bibinfo {author} {\bibfnamefont {Y.}~\bibnamefont {Wang}},
  \bibinfo {author} {\bibfnamefont {N.~D.}\ \bibnamefont {Patel}}, \bibinfo
  {author} {\bibfnamefont {G.}~\bibnamefont {Alvarez}}, \bibinfo {author}
  {\bibfnamefont {T.~A.}\ \bibnamefont {Maier}}, \bibinfo {author}
  {\bibfnamefont {E.}~\bibnamefont {Dagotto}},\ and\ \bibinfo {author}
  {\bibfnamefont {S.}~\bibnamefont {Johnston}},\ }\bibfield  {title} {\bibinfo
  {title} {Doping evolution of charge and spin excitations in two-leg hubbard
  ladders: Comparing dmrg and flex results},\ }\href
  {https://doi.org/10.1103/PhysRevB.97.195156} {\bibfield  {journal} {\bibinfo
  {journal} {Phys. Rev. B}\ }\textbf {\bibinfo {volume} {97}},\ \bibinfo
  {pages} {195156} (\bibinfo {year} {2018})}\BibitemShut {NoStop}%
\bibitem [{\citenamefont {Qin}\ \emph {et~al.}(2020)\citenamefont {Qin},
  \citenamefont {Chung}, \citenamefont {Shi}, \citenamefont {Vitali},
  \citenamefont {Hubig}, \citenamefont {Schollw\"ock}, \citenamefont {White},\
  and\ \citenamefont {Zhang}}]{qin_etal_prx}%
  \BibitemOpen
  \bibfield  {author} {\bibinfo {author} {\bibfnamefont {M.}~\bibnamefont
  {Qin}}, \bibinfo {author} {\bibfnamefont {C.-M.}\ \bibnamefont {Chung}},
  \bibinfo {author} {\bibfnamefont {H.}~\bibnamefont {Shi}}, \bibinfo {author}
  {\bibfnamefont {E.}~\bibnamefont {Vitali}}, \bibinfo {author} {\bibfnamefont
  {C.}~\bibnamefont {Hubig}}, \bibinfo {author} {\bibfnamefont
  {U.}~\bibnamefont {Schollw\"ock}}, \bibinfo {author} {\bibfnamefont {S.~R.}\
  \bibnamefont {White}},\ and\ \bibinfo {author} {\bibfnamefont
  {S.}~\bibnamefont {Zhang}} (\bibinfo {collaboration} {Simons Collaboration on
  the Many-Electron Problem}),\ }\bibfield  {title} {\bibinfo {title} {Absence
  of superconductivity in the pure two-dimensional hubbard model},\ }\href
  {https://doi.org/10.1103/PhysRevX.10.031016} {\bibfield  {journal} {\bibinfo
  {journal} {Phys. Rev. X}\ }\textbf {\bibinfo {volume} {10}},\ \bibinfo
  {pages} {031016} (\bibinfo {year} {2020})}\BibitemShut {NoStop}%
\bibitem [{\citenamefont {Kinnunen}\ \emph {et~al.}(2018)\citenamefont
  {Kinnunen}, \citenamefont {Baarsma}, \citenamefont {Martikainen},\ and\
  \citenamefont {Törmä}}]{kinnunen_etal_rpp_18}%
  \BibitemOpen
  \bibfield  {author} {\bibinfo {author} {\bibfnamefont {J.~J.}\ \bibnamefont
  {Kinnunen}}, \bibinfo {author} {\bibfnamefont {J.~E.}\ \bibnamefont
  {Baarsma}}, \bibinfo {author} {\bibfnamefont {J.-P.}\ \bibnamefont
  {Martikainen}},\ and\ \bibinfo {author} {\bibfnamefont {P.}~\bibnamefont
  {Törmä}},\ }\bibfield  {title} {\bibinfo {title} {The
  fulde–ferrell–larkin–ovchinnikov state for ultracold fermions in
  lattice and harmonic potentials: a review},\ }\href
  {https://doi.org/10.1088/1361-6633/aaa4ad} {\bibfield  {journal} {\bibinfo
  {journal} {Reports on Progress in Physics}\ }\textbf {\bibinfo {volume}
  {81}},\ \bibinfo {pages} {046401} (\bibinfo {year} {2018})}\BibitemShut
  {NoStop}%
\bibitem [{\citenamefont {Baarsma}\ and\ \citenamefont
  {Törmä}(2016)}]{baarsma_jmo_16}%
  \BibitemOpen
  \bibfield  {author} {\bibinfo {author} {\bibfnamefont {J.~E.}\ \bibnamefont
  {Baarsma}}\ and\ \bibinfo {author} {\bibfnamefont {P.}~\bibnamefont
  {Törmä}},\ }\bibfield  {title} {\bibinfo {title} {Larkin-ovchinnikov phases
  in two-dimensional square lattices},\ }\href@noop {} {\bibfield  {journal}
  {\bibinfo  {journal} {Journal of Modern Optics}\ }\textbf {\bibinfo {volume}
  {63}},\ \bibinfo {pages} {1795} (\bibinfo {year} {2016})}\BibitemShut
  {NoStop}%
\bibitem [{\citenamefont {Baarsma}\ and\ \citenamefont
  {Stoof}(2013)}]{baarsma_stoof_pra_13}%
  \BibitemOpen
  \bibfield  {author} {\bibinfo {author} {\bibfnamefont {J.~E.}\ \bibnamefont
  {Baarsma}}\ and\ \bibinfo {author} {\bibfnamefont {H.~T.~C.}\ \bibnamefont
  {Stoof}},\ }\bibfield  {title} {\bibinfo {title} {Inhomogeneous superfluid
  phases in ${}^{6}$li-${}^{40}$k mixtures at unitarity},\ }\href
  {https://doi.org/10.1103/PhysRevA.87.063612} {\bibfield  {journal} {\bibinfo
  {journal} {Phys. Rev. A}\ }\textbf {\bibinfo {volume} {87}},\ \bibinfo
  {pages} {063612} (\bibinfo {year} {2013})}\BibitemShut {NoStop}%
\bibitem [{\citenamefont {Mora}\ and\ \citenamefont
  {Combescot}(2005)}]{mora_combescot_05}%
  \BibitemOpen
  \bibfield  {author} {\bibinfo {author} {\bibfnamefont {C.}~\bibnamefont
  {Mora}}\ and\ \bibinfo {author} {\bibfnamefont {R.}~\bibnamefont
  {Combescot}},\ }\bibfield  {title} {\bibinfo {title} {Transition to
  fulde-ferrell-larkin-ovchinnikov phases in three dimensions: A quasiclassical
  investigation at low temperature with fourier expansion},\ }\href
  {https://doi.org/10.1103/PhysRevB.71.214504} {\bibfield  {journal} {\bibinfo
  {journal} {Phys. Rev. B}\ }\textbf {\bibinfo {volume} {71}},\ \bibinfo
  {pages} {214504} (\bibinfo {year} {2005})}\BibitemShut {NoStop}%
\bibitem [{\citenamefont {Yoshida}\ and\ \citenamefont
  {Yip}(2007)}]{yoshida_yip_pra_07}%
  \BibitemOpen
  \bibfield  {author} {\bibinfo {author} {\bibfnamefont {N.}~\bibnamefont
  {Yoshida}}\ and\ \bibinfo {author} {\bibfnamefont {S.-K.}\ \bibnamefont
  {Yip}},\ }\bibfield  {title} {\bibinfo {title} {Larkin-ovchinnikov state in
  resonant fermi gas},\ }\href {https://doi.org/10.1103/PhysRevA.75.063601}
  {\bibfield  {journal} {\bibinfo  {journal} {Phys. Rev. A}\ }\textbf {\bibinfo
  {volume} {75}},\ \bibinfo {pages} {063601} (\bibinfo {year}
  {2007})}\BibitemShut {NoStop}%
\bibitem [{\citenamefont {Schmid}\ \emph {et~al.}(2002)\citenamefont {Schmid},
  \citenamefont {Todo}, \citenamefont {Troyer},\ and\ \citenamefont
  {Dorneich}}]{schmid_etal_prl_02}%
  \BibitemOpen
  \bibfield  {author} {\bibinfo {author} {\bibfnamefont {G.}~\bibnamefont
  {Schmid}}, \bibinfo {author} {\bibfnamefont {S.}~\bibnamefont {Todo}},
  \bibinfo {author} {\bibfnamefont {M.}~\bibnamefont {Troyer}},\ and\ \bibinfo
  {author} {\bibfnamefont {A.}~\bibnamefont {Dorneich}},\ }\bibfield  {title}
  {\bibinfo {title} {Finite-temperature phase diagram of hard-core bosons in
  two dimensions},\ }\href {https://doi.org/10.1103/PhysRevLett.88.167208}
  {\bibfield  {journal} {\bibinfo  {journal} {Phys. Rev. Lett.}\ }\textbf
  {\bibinfo {volume} {88}},\ \bibinfo {pages} {167208} (\bibinfo {year}
  {2002})}\BibitemShut {NoStop}%
\bibitem [{\citenamefont {Gresista}\ \emph {et~al.}(2023)\citenamefont
  {Gresista}, \citenamefont {Kiese}, \citenamefont {Trebst},\ and\
  \citenamefont {Scherer}}]{Gresista_2023}%
  \BibitemOpen
  \bibfield  {author} {\bibinfo {author} {\bibfnamefont {L.}~\bibnamefont
  {Gresista}}, \bibinfo {author} {\bibfnamefont {D.}~\bibnamefont {Kiese}},
  \bibinfo {author} {\bibfnamefont {S.}~\bibnamefont {Trebst}},\ and\ \bibinfo
  {author} {\bibfnamefont {M.~M.}\ \bibnamefont {Scherer}},\ }\bibfield
  {title} {\bibinfo {title} {Spin-valley magnetism on the triangular moiré
  lattice with su(4) breaking interactions},\ }\bibfield  {journal} {\bibinfo
  {journal} {Physical Review B}\ }\textbf {\bibinfo {volume} {108}},\ \href
  {https://doi.org/10.1103/physrevb.108.045102} {10.1103/physrevb.108.045102}
  (\bibinfo {year} {2023})\BibitemShut {NoStop}%
\bibitem [{\citenamefont {Cao}\ \emph {et~al.}(2018)\citenamefont {Cao},
  \citenamefont {Fatemi}, \citenamefont {Fang}, \citenamefont {Watanabe},
  \citenamefont {Taniguchi}, \citenamefont {Kaxiras},\ and\ \citenamefont
  {Jarillo-Herrero}}]{cao_tbg_2018}%
  \BibitemOpen
  \bibfield  {author} {\bibinfo {author} {\bibfnamefont {Y.}~\bibnamefont
  {Cao}}, \bibinfo {author} {\bibfnamefont {V.}~\bibnamefont {Fatemi}},
  \bibinfo {author} {\bibfnamefont {S.}~\bibnamefont {Fang}}, \bibinfo {author}
  {\bibfnamefont {K.}~\bibnamefont {Watanabe}}, \bibinfo {author}
  {\bibfnamefont {T.}~\bibnamefont {Taniguchi}}, \bibinfo {author}
  {\bibfnamefont {E.}~\bibnamefont {Kaxiras}},\ and\ \bibinfo {author}
  {\bibfnamefont {P.}~\bibnamefont {Jarillo-Herrero}},\ }\bibfield  {title}
  {\bibinfo {title} {Unconventional superconductivity in magic-angle graphene
  superlattices},\ }\href@noop {} {\bibfield  {journal} {\bibinfo  {journal}
  {Nature}\ }\textbf {\bibinfo {volume} {556}},\ \bibinfo {pages} {43}
  (\bibinfo {year} {2018})}\BibitemShut {NoStop}%
\bibitem [{\citenamefont {Xia}\ \emph {et~al.}(2024)\citenamefont {Xia},
  \citenamefont {Han}, \citenamefont {Watanabe}, \citenamefont {Taniguchi},
  \citenamefont {Shan},\ and\ \citenamefont {Mak}}]{xia_2024}%
  \BibitemOpen
  \bibfield  {author} {\bibinfo {author} {\bibfnamefont {Y.}~\bibnamefont
  {Xia}}, \bibinfo {author} {\bibfnamefont {Z.}~\bibnamefont {Han}}, \bibinfo
  {author} {\bibfnamefont {K.}~\bibnamefont {Watanabe}}, \bibinfo {author}
  {\bibfnamefont {T.}~\bibnamefont {Taniguchi}}, \bibinfo {author}
  {\bibfnamefont {J.}~\bibnamefont {Shan}},\ and\ \bibinfo {author}
  {\bibfnamefont {K.~F.}\ \bibnamefont {Mak}},\ }\bibfield  {title} {\bibinfo
  {title} {Unconventional superconductivity in twisted bilayer wse2},\
  }\href@noop {} {\bibfield  {journal} {\bibinfo  {journal} {arXiv:2405.14784}\
  } (\bibinfo {year} {2024})}\BibitemShut {NoStop}%
\bibitem [{\citenamefont {Mora}\ \emph {et~al.}(2019)\citenamefont {Mora},
  \citenamefont {Regnault},\ and\ \citenamefont {Bernevig}}]{mora_etal_prl_19}%
  \BibitemOpen
  \bibfield  {author} {\bibinfo {author} {\bibfnamefont {C.}~\bibnamefont
  {Mora}}, \bibinfo {author} {\bibfnamefont {N.}~\bibnamefont {Regnault}},\
  and\ \bibinfo {author} {\bibfnamefont {B.~A.}\ \bibnamefont {Bernevig}},\
  }\bibfield  {title} {\bibinfo {title} {Flatbands and perfect metal in
  trilayer moir\'e graphene},\ }\href
  {https://doi.org/10.1103/PhysRevLett.123.026402} {\bibfield  {journal}
  {\bibinfo  {journal} {Phys. Rev. Lett.}\ }\textbf {\bibinfo {volume} {123}},\
  \bibinfo {pages} {026402} (\bibinfo {year} {2019})}\BibitemShut {NoStop}%
\bibitem [{\citenamefont {Khalaf}\ \emph {et~al.}(2019)\citenamefont {Khalaf},
  \citenamefont {Kruchkov}, \citenamefont {Tarnopolsky},\ and\ \citenamefont
  {Vishwanath}}]{khalaf_etal_prb_19}%
  \BibitemOpen
  \bibfield  {author} {\bibinfo {author} {\bibfnamefont {E.}~\bibnamefont
  {Khalaf}}, \bibinfo {author} {\bibfnamefont {A.~J.}\ \bibnamefont
  {Kruchkov}}, \bibinfo {author} {\bibfnamefont {G.}~\bibnamefont
  {Tarnopolsky}},\ and\ \bibinfo {author} {\bibfnamefont {A.}~\bibnamefont
  {Vishwanath}},\ }\bibfield  {title} {\bibinfo {title} {Magic angle hierarchy
  in twisted graphene multilayers},\ }\href
  {https://doi.org/10.1103/PhysRevB.100.085109} {\bibfield  {journal} {\bibinfo
   {journal} {Phys. Rev. B}\ }\textbf {\bibinfo {volume} {100}},\ \bibinfo
  {pages} {085109} (\bibinfo {year} {2019})}\BibitemShut {NoStop}%
\bibitem [{\citenamefont {Carr}\ \emph {et~al.}(2020)\citenamefont {Carr},
  \citenamefont {Li}, \citenamefont {Zhu}, \citenamefont {Kaxiras},
  \citenamefont {Sachdev},\ and\ \citenamefont {Kruchkov}}]{carr_etal_nl_20}%
  \BibitemOpen
  \bibfield  {author} {\bibinfo {author} {\bibfnamefont {S.}~\bibnamefont
  {Carr}}, \bibinfo {author} {\bibfnamefont {C.}~\bibnamefont {Li}}, \bibinfo
  {author} {\bibfnamefont {Z.}~\bibnamefont {Zhu}}, \bibinfo {author}
  {\bibfnamefont {E.}~\bibnamefont {Kaxiras}}, \bibinfo {author} {\bibfnamefont
  {S.}~\bibnamefont {Sachdev}},\ and\ \bibinfo {author} {\bibfnamefont
  {A.}~\bibnamefont {Kruchkov}},\ }\bibfield  {title} {\bibinfo {title}
  {Ultraheavy and ultrarelativistic dirac quasiparticles in sandwiched
  graphenes},\ }\href {https://doi.org/10.1021/acs.nanolett.9b04979} {\bibfield
   {journal} {\bibinfo  {journal} {Nano Letters}\ }\textbf {\bibinfo {volume}
  {20}},\ \bibinfo {pages} {3030} (\bibinfo {year} {2020})},\ \bibinfo {note}
  {pMID: 32208724},\ \Eprint
  {https://arxiv.org/abs/https://doi.org/10.1021/acs.nanolett.9b04979}
  {https://doi.org/10.1021/acs.nanolett.9b04979} \BibitemShut {NoStop}%
\bibitem [{\citenamefont {Lei}\ \emph {et~al.}(2021)\citenamefont {Lei},
  \citenamefont {Linhart}, \citenamefont {Qin}, \citenamefont {Libisch},\ and\
  \citenamefont {MacDonald}}]{lei_etal_prb_21}%
  \BibitemOpen
  \bibfield  {author} {\bibinfo {author} {\bibfnamefont {C.}~\bibnamefont
  {Lei}}, \bibinfo {author} {\bibfnamefont {L.}~\bibnamefont {Linhart}},
  \bibinfo {author} {\bibfnamefont {W.}~\bibnamefont {Qin}}, \bibinfo {author}
  {\bibfnamefont {F.}~\bibnamefont {Libisch}},\ and\ \bibinfo {author}
  {\bibfnamefont {A.~H.}\ \bibnamefont {MacDonald}},\ }\bibfield  {title}
  {\bibinfo {title} {Mirror symmetry breaking and lateral stacking shifts in
  twisted trilayer graphene},\ }\href
  {https://doi.org/10.1103/PhysRevB.104.035139} {\bibfield  {journal} {\bibinfo
   {journal} {Phys. Rev. B}\ }\textbf {\bibinfo {volume} {104}},\ \bibinfo
  {pages} {035139} (\bibinfo {year} {2021})}\BibitemShut {NoStop}%
\bibitem [{\citenamefont {C\ifmmode \u{a}\else \u{a}\fi{}lug\ifmmode~\u{a}\else
  \u{a}\fi{}ru}\ \emph {et~al.}(2021)\citenamefont {C\ifmmode \u{a}\else
  \u{a}\fi{}lug\ifmmode~\u{a}\else \u{a}\fi{}ru}, \citenamefont {Xie},
  \citenamefont {Song}, \citenamefont {Lian}, \citenamefont {Regnault},\ and\
  \citenamefont {Bernevig}}]{calugaru_etal_prb_21}%
  \BibitemOpen
  \bibfield  {author} {\bibinfo {author} {\bibfnamefont {D.}~\bibnamefont
  {C\ifmmode \u{a}\else \u{a}\fi{}lug\ifmmode~\u{a}\else \u{a}\fi{}ru}},
  \bibinfo {author} {\bibfnamefont {F.}~\bibnamefont {Xie}}, \bibinfo {author}
  {\bibfnamefont {Z.-D.}\ \bibnamefont {Song}}, \bibinfo {author}
  {\bibfnamefont {B.}~\bibnamefont {Lian}}, \bibinfo {author} {\bibfnamefont
  {N.}~\bibnamefont {Regnault}},\ and\ \bibinfo {author} {\bibfnamefont
  {B.~A.}\ \bibnamefont {Bernevig}},\ }\bibfield  {title} {\bibinfo {title}
  {Twisted symmetric trilayer graphene: Single-particle and many-body
  hamiltonians and hidden nonlocal symmetries of trilayer moir\'e systems with
  and without displacement field},\ }\href
  {https://doi.org/10.1103/PhysRevB.103.195411} {\bibfield  {journal} {\bibinfo
   {journal} {Phys. Rev. B}\ }\textbf {\bibinfo {volume} {103}},\ \bibinfo
  {pages} {195411} (\bibinfo {year} {2021})}\BibitemShut {NoStop}%
\bibitem [{\citenamefont {Shin}\ \emph {et~al.}(2021)\citenamefont {Shin},
  \citenamefont {Chittari},\ and\ \citenamefont {Jung}}]{shin_etal_prb_21}%
  \BibitemOpen
  \bibfield  {author} {\bibinfo {author} {\bibfnamefont {J.}~\bibnamefont
  {Shin}}, \bibinfo {author} {\bibfnamefont {B.~L.}\ \bibnamefont {Chittari}},\
  and\ \bibinfo {author} {\bibfnamefont {J.}~\bibnamefont {Jung}},\ }\bibfield
  {title} {\bibinfo {title} {Stacking and gate-tunable topological flat bands,
  gaps, and anisotropic strip patterns in twisted trilayer graphene},\ }\href
  {https://doi.org/10.1103/PhysRevB.104.045413} {\bibfield  {journal} {\bibinfo
   {journal} {Phys. Rev. B}\ }\textbf {\bibinfo {volume} {104}},\ \bibinfo
  {pages} {045413} (\bibinfo {year} {2021})}\BibitemShut {NoStop}%
\bibitem [{\citenamefont {Hao}\ \emph {et~al.}(2021)\citenamefont {Hao},
  \citenamefont {Zimmerman}, \citenamefont {Ledwith}, \citenamefont {Khalaf},
  \citenamefont {Najafabadi}, \citenamefont {Watanabe}, \citenamefont
  {Taniguchi}, \citenamefont {Vishwanath},\ and\ \citenamefont
  {Kim}}]{hao_etal_science_21}%
  \BibitemOpen
  \bibfield  {author} {\bibinfo {author} {\bibfnamefont {Z.}~\bibnamefont
  {Hao}}, \bibinfo {author} {\bibfnamefont {A.~M.}\ \bibnamefont {Zimmerman}},
  \bibinfo {author} {\bibfnamefont {P.}~\bibnamefont {Ledwith}}, \bibinfo
  {author} {\bibfnamefont {E.}~\bibnamefont {Khalaf}}, \bibinfo {author}
  {\bibfnamefont {D.~H.}\ \bibnamefont {Najafabadi}}, \bibinfo {author}
  {\bibfnamefont {K.}~\bibnamefont {Watanabe}}, \bibinfo {author}
  {\bibfnamefont {T.}~\bibnamefont {Taniguchi}}, \bibinfo {author}
  {\bibfnamefont {A.}~\bibnamefont {Vishwanath}},\ and\ \bibinfo {author}
  {\bibfnamefont {P.}~\bibnamefont {Kim}},\ }\bibfield  {title} {\bibinfo
  {title} {Electric field\&\#x2013;tunable superconductivity in
  alternating-twist magic-angle trilayer graphene},\ }\href
  {https://doi.org/10.1126/science.abg0399} {\bibfield  {journal} {\bibinfo
  {journal} {Science}\ }\textbf {\bibinfo {volume} {371}},\ \bibinfo {pages}
  {1133} (\bibinfo {year} {2021})},\ \Eprint
  {https://arxiv.org/abs/https://www.science.org/doi/pdf/10.1126/science.abg0399}
  {https://www.science.org/doi/pdf/10.1126/science.abg0399} \BibitemShut
  {NoStop}%
\bibitem [{\citenamefont {Christos}\ \emph {et~al.}(2022)\citenamefont
  {Christos}, \citenamefont {Sachdev},\ and\ \citenamefont
  {Scheurer}}]{christos_etal_prx_22}%
  \BibitemOpen
  \bibfield  {author} {\bibinfo {author} {\bibfnamefont {M.}~\bibnamefont
  {Christos}}, \bibinfo {author} {\bibfnamefont {S.}~\bibnamefont {Sachdev}},\
  and\ \bibinfo {author} {\bibfnamefont {M.~S.}\ \bibnamefont {Scheurer}},\
  }\bibfield  {title} {\bibinfo {title} {Correlated insulators, semimetals, and
  superconductivity in twisted trilayer graphene},\ }\href
  {https://doi.org/10.1103/PhysRevX.12.021018} {\bibfield  {journal} {\bibinfo
  {journal} {Phys. Rev. X}\ }\textbf {\bibinfo {volume} {12}},\ \bibinfo
  {pages} {021018} (\bibinfo {year} {2022})}\BibitemShut {NoStop}%
\bibitem [{\citenamefont {Fischer}\ \emph {et~al.}(2022)\citenamefont
  {Fischer}, \citenamefont {Goodwin}, \citenamefont {Mostofi}, \citenamefont
  {Lischner}, \citenamefont {Kennes},\ and\ \citenamefont
  {Klebl}}]{fischer_etal_nat_22}%
  \BibitemOpen
  \bibfield  {author} {\bibinfo {author} {\bibfnamefont {A.}~\bibnamefont
  {Fischer}}, \bibinfo {author} {\bibfnamefont {Z.~A.~H.}\ \bibnamefont
  {Goodwin}}, \bibinfo {author} {\bibfnamefont {A.~A.}\ \bibnamefont
  {Mostofi}}, \bibinfo {author} {\bibfnamefont {J.}~\bibnamefont {Lischner}},
  \bibinfo {author} {\bibfnamefont {D.~M.}\ \bibnamefont {Kennes}},\ and\
  \bibinfo {author} {\bibfnamefont {L.}~\bibnamefont {Klebl}},\ }\bibfield
  {title} {\bibinfo {title} {Unconventional superconductivity in magic-angle
  twisted trilayer graphene},\ }\href
  {https://doi.org/10.1038/s41535-021-00410-w} {\bibfield  {journal} {\bibinfo
  {journal} {npj Quantum Materials}\ }\textbf {\bibinfo {volume} {7}},\
  \bibinfo {pages} {5} (\bibinfo {year} {2022})}\BibitemShut {NoStop}%
\bibitem [{\citenamefont {Wang}\ and\ \citenamefont
  {Vishwanath}(2009)}]{PhysRevB.80.064413}%
  \BibitemOpen
  \bibfield  {author} {\bibinfo {author} {\bibfnamefont {F.}~\bibnamefont
  {Wang}}\ and\ \bibinfo {author} {\bibfnamefont {A.}~\bibnamefont
  {Vishwanath}},\ }\bibfield  {title} {\bibinfo {title} {${\text{z}}_{2}$
  spin-orbital liquid state in the square lattice kugel-khomskii model},\
  }\href {https://doi.org/10.1103/PhysRevB.80.064413} {\bibfield  {journal}
  {\bibinfo  {journal} {Phys. Rev. B}\ }\textbf {\bibinfo {volume} {80}},\
  \bibinfo {pages} {064413} (\bibinfo {year} {2009})}\BibitemShut {NoStop}%
\bibitem [{\citenamefont {Corboz}\ \emph {et~al.}(2012)\citenamefont {Corboz},
  \citenamefont {Lajk\'o}, \citenamefont {L\"auchli}, \citenamefont {Penc},\
  and\ \citenamefont {Mila}}]{PhysRevX.2.041013}%
  \BibitemOpen
  \bibfield  {author} {\bibinfo {author} {\bibfnamefont {P.}~\bibnamefont
  {Corboz}}, \bibinfo {author} {\bibfnamefont {M.}~\bibnamefont {Lajk\'o}},
  \bibinfo {author} {\bibfnamefont {A.~M.}\ \bibnamefont {L\"auchli}}, \bibinfo
  {author} {\bibfnamefont {K.}~\bibnamefont {Penc}},\ and\ \bibinfo {author}
  {\bibfnamefont {F.}~\bibnamefont {Mila}},\ }\bibfield  {title} {\bibinfo
  {title} {Spin-orbital quantum liquid on the honeycomb lattice},\ }\href
  {https://doi.org/10.1103/PhysRevX.2.041013} {\bibfield  {journal} {\bibinfo
  {journal} {Phys. Rev. X}\ }\textbf {\bibinfo {volume} {2}},\ \bibinfo {pages}
  {041013} (\bibinfo {year} {2012})}\BibitemShut {NoStop}%
\bibitem [{\citenamefont {{J\"{u}lich Supercomputing Centre}}(2021)}]{JURECA}%
  \BibitemOpen
  \bibfield  {author} {\bibinfo {author} {\bibnamefont {{J\"{u}lich
  Supercomputing Centre}}},\ }\bibfield  {title} {\bibinfo {title} {{JURECA:
  Data Centric and Booster Modules implementing the Modular Supercomputing
  Architecture at J\"{u}lich Supercomputing Centre}},\ }\bibfield  {journal}
  {\bibinfo  {journal} {Journal of large-scale research facilities}\ }\textbf
  {\bibinfo {volume} {7}},\ \href {https://doi.org/10.17815/jlsrf-7-182}
  {10.17815/jlsrf-7-182} (\bibinfo {year} {2021})\BibitemShut {NoStop}%
\bibitem [{\citenamefont {Hauschild}\ and\ \citenamefont
  {Pollmann}(2018)}]{tenpy}%
  \BibitemOpen
  \bibfield  {author} {\bibinfo {author} {\bibfnamefont {J.}~\bibnamefont
  {Hauschild}}\ and\ \bibinfo {author} {\bibfnamefont {F.}~\bibnamefont
  {Pollmann}},\ }\bibfield  {title} {\bibinfo {title} {{Efficient numerical
  simulations with Tensor Networks: Tensor Network Python (TeNPy)}},\ }\href
  {https://doi.org/10.21468/SciPostPhysLectNotes.5} {\bibfield  {journal}
  {\bibinfo  {journal} {SciPost Phys. Lect. Notes}\ ,\ \bibinfo {pages} {5}}
  (\bibinfo {year} {2018})},\ \bibinfo {note} {code available from
  \url{https://github.com/tenpy/tenpy}},\ \Eprint
  {https://arxiv.org/abs/1805.00055} {arXiv:1805.00055} \BibitemShut {NoStop}%
\bibitem [{\citenamefont {Dolfi}\ \emph {et~al.}(2015)\citenamefont {Dolfi},
  \citenamefont {Bauer}, \citenamefont {Keller},\ and\ \citenamefont
  {Troyer}}]{dolfi_etal_prb_15}%
  \BibitemOpen
  \bibfield  {author} {\bibinfo {author} {\bibfnamefont {M.}~\bibnamefont
  {Dolfi}}, \bibinfo {author} {\bibfnamefont {B.}~\bibnamefont {Bauer}},
  \bibinfo {author} {\bibfnamefont {S.}~\bibnamefont {Keller}},\ and\ \bibinfo
  {author} {\bibfnamefont {M.}~\bibnamefont {Troyer}},\ }\bibfield  {title}
  {\bibinfo {title} {Pair correlations in doped hubbard ladders},\ }\href
  {https://doi.org/10.1103/PhysRevB.92.195139} {\bibfield  {journal} {\bibinfo
  {journal} {Phys. Rev. B}\ }\textbf {\bibinfo {volume} {92}},\ \bibinfo
  {pages} {195139} (\bibinfo {year} {2015})}\BibitemShut {NoStop}%
\bibitem [{\citenamefont {Hubig}\ \emph {et~al.}(2018)\citenamefont {Hubig},
  \citenamefont {Haegeman},\ and\ \citenamefont
  {Schollw\"ock}}]{hubig_etal_prb_18}%
  \BibitemOpen
  \bibfield  {author} {\bibinfo {author} {\bibfnamefont {C.}~\bibnamefont
  {Hubig}}, \bibinfo {author} {\bibfnamefont {J.}~\bibnamefont {Haegeman}},\
  and\ \bibinfo {author} {\bibfnamefont {U.}~\bibnamefont {Schollw\"ock}},\
  }\bibfield  {title} {\bibinfo {title} {Error estimates for extrapolations
  with matrix-product states},\ }\href
  {https://doi.org/10.1103/PhysRevB.97.045125} {\bibfield  {journal} {\bibinfo
  {journal} {Phys. Rev. B}\ }\textbf {\bibinfo {volume} {97}},\ \bibinfo
  {pages} {045125} (\bibinfo {year} {2018})}\BibitemShut {NoStop}%
\end{thebibliography}%

\end{document}